\documentclass[fleqn,usenatbib]{mnras}

\usepackage{newtxtext,newtxmath,cuted,lipsum}
\usepackage[T1]{fontenc} 
\usepackage{graphicx}	
\usepackage{amsmath}	
\usepackage{amsfonts}
\usepackage{enumerate} 
\usepackage{colordvi} 
\usepackage{bm}
\usepackage{epsfig}
\usepackage{float}
\usepackage{verbatim}
\usepackage{subfig}
\usepackage[dvipsnames]{xcolor}
\usepackage{multirow,multicol}

\usepackage[colorinlistoftodos]{todonotes}
\usepackage{hyperref}
\usepackage{ulem}

\usepackage[colorinlistoftodos]{todonotes}
\usepackage{hyperref}

\newcommand{\chh}{\color{black}}

\newcommand{\B}[1]{\textcolor{black}{#1}} 

\definecolor{darkgreen}{cmyk}{0.85,0.2,1.00,0.05}

\definecolor{pink}{rgb}{0.85, 0.2, 0.53}

\usepackage{tikz}
\usetikzlibrary{shapes.geometric, arrows}

\tikzstyle{start} = [trapezium, trapezium left angle=70, trapezium right angle=110, minimum width=0.5cm, minimum height=0.5cm, text centered, draw=black, fill=orange!60]

\tikzstyle{start2} = [trapezium, trapezium left angle=70, trapezium right angle=110, minimum width=0.5cm, minimum height=0.5cm, text centered, draw=black, fill=yellow!60]

\tikzstyle{io_halo} = [rectangle, trapezium left angle=70, trapezium right angle=110, minimum width=0.5cm, minimum height=0.5cm, text centered, draw=black, fill=red!50]

\tikzstyle{io_halo2} = [rectangle, trapezium left angle=70, trapezium right angle=110, minimum width=0.5cm, minimum height=0.5cm, text centered, draw=black, fill=cyan!50]

\tikzstyle{io_pseudo} = [rectangle, trapezium left angle=70, trapezium right angle=110, minimum width=2cm, minimum height=1cm, text centered, draw=black, fill=gray!40]

\tikzstyle{reaction} = [rectangle, trapezium left angle=70, trapezium right angle=110, minimum width=2cm, minimum height=1cm, text centered, draw=black, fill=blue!40]

\tikzstyle{io_nl_nofeed} = [trapezium, trapezium left angle=70, trapezium right angle=110, minimum height=1cm, text centered, draw=black, fill=green!40]

\tikzstyle{io_feed} = [trapezium, trapezium left angle=70, trapezium right angle=110, minimum height=1cm, text centered, draw=black, fill=yellow!30]

\tikzstyle{io_nl} = [trapezium, trapezium left angle=70, trapezium right angle=110, minimum height=1cm, text centered, draw=black, fill=green!80]

\tikzstyle{arrow} = [thick,->,>=stealth]

\title[Non-linear reaction beyond $\Lambda$CDM with massive neutrinos]{On the road to percent accuracy V: the non-linear power spectrum beyond $\Lambda$CDM with massive neutrinos and baryonic feedback}  

\author[B. Bose et al.]{Benjamin Bose$^{1}$\thanks{E-mail:benjamin.bose@unige.ch},
Bill S. Wright$^{2}$, 
Matteo Cataneo$^{3}$, 
Alkistis Pourtsidou$^{2,4}$, 
Carlo Giocoli$^{5,6}$,
\newauthor
Lucas Lombriser$^{1}$,
Ian G. McCarthy$^{7}$,
Marco Baldi$^{8,5,6}$,
Simon Pfeifer$^{9}$,
Qianli Xia$^{3}$.
\\
$^{1}$D\'epartement de Physique Th\'eorique, Universit\'e de Gen\`eve, 24 quai Ernest Ansermet, 1211 Gen\`eve 4, Switzerland. \\
$^{2}$ School of Physics and Astronomy, Queen Mary University of London, Mile End Road, London E1 4NS, U.K. \\
$^{3}$Institute for Astronomy, University of Edinburgh, Royal Observatory, Blackford Hill, Edinburgh, EH9 3HJ, U.K. \\
$^{4}$Department of Physics \& Astronomy, University of the Western Cape, Cape Town 7535, South Africa. \\
$^{5}$INAF - Osservatorio di Astrofisica e Scienza dello Spazio di Bologna, via Gobetti 93/3, I-40129 Bologna, Italy. \\
$^{6}$INFN - Sezione di Bologna, viale Berti Pichat 6/2, I-40127 Bologna, Italy. \\ 
$^{7}$ Astrophysics Research Institute, Liverpool John Moores University, 146 Brownlow Hill, Liverpool L3 5RF, UK.\\
$^{8}$Dipartimento di Fisica e Astronomia "Augusto Righi", Alma Mater Studiorum Universit\`{a} di Bologna, via Gobetti 93/2, I-40129 Bologna, Italy.\\
$^{9}$Leibniz-Institut für Astrophysik Potsdam, An der Sternwarte 16, D-14482 Potsdam, Germany. 
}

\date{Accepted XXX. Received YYY; in original form ZZZ}

\pubyear{2021}

\begin{document}
\label{firstpage}
\pagerange{\pageref{firstpage}--\pageref{lastpage}}
\maketitle

\begin{abstract}
In the context of forthcoming galaxy surveys, to ensure unbiased constraints on cosmology and gravity when using non-linear structure information, percent-level accuracy is required when modelling the power spectrum. This calls for frameworks that can accurately capture the relevant physical effects, while allowing for deviations from $\Lambda$CDM. Massive neutrino and baryonic physics are two of the most relevant such effects. We present an integration of the halo model reaction frameworks for massive neutrinos and beyond-$\Lambda$CDM cosmologies. The integrated halo model reaction, combined with a pseudo power spectrum modelled by {\tt HMCode2020} is then compared against $N$-body simulations that include both massive neutrinos and an $f(R)$ modification to gravity. We find that the framework is 4\% accurate down to at least $k\approx 3 \, h/{\rm Mpc}$ for a modification to gravity of $|f_{\rm R0}|\leq 10^{-5}$  and for the total neutrino mass $M_\nu \equiv \sum m_\nu \leq 0.15$ eV. We also find that the framework is 4\% consistent with EuclidEmulator2 as well as the Bacco emulator for \B{most of the considered} $\nu w$CDM cosmologies down to at least $k \approx 3 \, h$/Mpc. Finally, we compare against hydrodynamical simulations employing {\tt HMCode2020}'s baryonic feedback modelling on top of the halo model reaction. For $\nu \Lambda$CDM cosmologies we find 2\% accuracy for $M_\nu \leq 0.48$eV down to at least $k\approx 5h$/Mpc. Similar accuracy is found when comparing to $\nu w$CDM  hydrodynamical simulations with $M_\nu = 0.06$eV. This offers the first non-linear, theoretically general means of accurately including massive neutrinos for beyond-$\Lambda$CDM cosmologies, and further suggests that baryonic, massive neutrino and dark energy physics can be reliably modelled independently. 
\end{abstract}


\section{Introduction}
The standard model of cosmology, $\Lambda$CDM, is extraordinarily consistent with a wealth of cosmological data sets, from the cosmic microwave background measurements \citep[CMB,][]{Aghanim:2018eyx} to measurements of the large-scale structure of the Universe \citep[LSS,][]{Anderson:2012sa,Song:2015oza,Beutler:2016arn,Hildebrandt:2016iqg,2021A&A...646A.140H,Abbott:2020knk}. Despite this success, the model comes with the highly contentious perquisite that $95\%$ of the matter-energy content of the Universe today is `dark', i.e.  which have so far have not been directly detected - cold dark matter (CDM) and a constant dark energy ($\Lambda$). Without understatement, this so called `dark sector' is one of the biggest problems in theoretical physics. 

In order to gain insight into this problem, the underlying assumptions of $\Lambda$CDM should be tested. Two of these key assumptions are:
\begin{itemize}
    \item 
    Dark energy is non-evolving. 
    \item 
    General relativity (GR) is applicable at all scales. 
\end{itemize}
Various alternatives to these assumptions have been proposed, coming in the form of dynamical dark energy \citep[for reviews see][]{Copeland:2006wr,Li:2011sd} and modifications to gravity (MG) \citep[for reviews see][]{Clifton:2011jh,Joyce:2016vqv,Koyama:2018som}. Despite the vast theoretical space which has been developed, much of this has been very well constrained by cosmological observations \citep[for a review of recent constraints see][]{Ferreira:2019xrr,Huterer:2017buf,Noller:2020afd}. 

One regime where cosmological and gravitational models are yet to be stringently tested is at the non-linear scales of LSS. Forthcoming galaxy surveys promise minute statistical errors at these scales (Euclid \cite{Amendola:2016saw, Blanchard:2019oqi}, DESI \cite{Levi:2019ggs}, Nancy Grace Roman Space Telescope  \cite{2019arXiv190205569A}, Vera Rubin Observatory  \cite{Abate:2012za}) which enable the detection of even the tiniest deviations to the standard model. This all hinges on our ability to theoretically model the key observables at these scales, including deviations to the standard model, at the percent level \citep{Taylor:2018nrc}. 

The key quantity of interest when considering LSS observations is the 2-point correlation function, or power spectrum in Fourier space, of the cosmological matter density field. This quantity is sensitive at non-linear scales to a host of physical effects which add new layers of complexity on top of the gravitational and cosmological modelling. In particular, the effects of a non-zero neutrino mass have been shown to be significant at the scales of interest \citep{2018MNRAS.481.1486B,2012MNRAS.420.2551B,2014JCAP...11..039B,2016MNRAS.459.1468M,2017ApJ...847...50L,Tram:2018znz,Massara:2014kba,2020arXiv200406245A}. Further, baryonic processes also begin to play a role the further we go into the non-linear regime (e.g., \citealt{vanDaalen2011,Mummery2017,Springel2018,vanDaalen:2019pst}; for a review see \citealt{Chisari:2019tus}). If we do not account for these effects, we will not be able to reliably use  the precise non-linear information coming from future surveys. For example, using these scales without accounting for phenomena such as baryonic feedback has been shown to produce biased estimates of cosmological parameters in the context of surveys like Euclid \citep{Semboloni:2011fe,Schneider:2019snl, Martinelli:2020yto}. Therefore, there is a pressing need for good theoretical models of these effects to be integrated in accurate frameworks for the matter power spectrum in beyond-$\Lambda$CDM cosmologies. 

Recently, a framework called the {\it halo model reaction} was proposed \citep{Cataneo:2018cic} \B{(for a precursor see \cite{Mead:2016ybv})} which offers a means of calculating the non-linear matter power spectrum at percent level accuracy in models beyond $\Lambda$CDM. A subsequent code called {\tt ReACT} \citep{Bose:2020wch} was developed, providing a means to efficiently compute the halo model reaction, making the framework viable for statistical data analyses, with a first application to constrain modified gravity using weak lensing data being made in \cite{Troester:2020}. Moreover, in \cite{Cataneo:2019fjp}, the halo model reaction was developed for massive neutrino cosmologies, assuming GR and a constant dark energy. With respect to baryonic effects, a number of modelling approaches have been developed which are based on parametrising feedback processes and then fitting to hydrodynamical simulations \citep{Mead:2020vgs,Schneider:2019snl,Schneider:2019xpf,Arico:2020lhq}. These promising prescriptions are yet to be integrated and tested against $N$-body simulations that include multiple physical effects simultaneously. 

In this paper, we present an extension to the framework of \cite{Cataneo:2018cic} (C19) to include the effects of massive neutrinos as modelled in \cite{Cataneo:2019fjp}, i.e.  consistently combining the beyond-$\Lambda$CDM and massive neutrino halo model reactions. We also include these extensions in {\tt ReACT}\footnote{Download {\tt ReACT} with massive neutrinos: \url{https://github.com/nebblu/ReACT/tree/react_with_neutrinos}}, making fast and accurate predictions for the non-linear power spectrum in beyond-$\Lambda$CDM cosmologies including massive neutrinos. We test the modelling against $N$-body simulations in $f(R)$ gravity and against the recently developed EuclidEmulator2 \citep{Knabenhans:2020gdo}  and the Bacco emulator \citep{2020arXiv200406245A} for evolving dark energy cosmologies with massive neutrinos ($\nu w$CDM). Finally, we also check the accuracy of {\tt ReACT} combined with the baryonic feedback fit of \cite{Mead:2020vgs} against hydrodynamical simulations that include both massive neutrino effects in standard ($\nu \Lambda$CDM) and evolving dark energy cosmologies.

This paper is organised as follows: In \autoref{sec:theory} we present the halo model reaction framework used to compute general modifications to $\Lambda$CDM non-linear power spectra with the inclusion of massive neutrinos.  In \autoref{sec:results} we assess the halo model reaction's accuracy through $N$-body simulations, state-of-the-art emulators and hydrodynamical simulation comparisons. In \autoref{sec:summary} we summarise our results and conclude.


\section{Extended halo model reaction}\label{sec:theory}

Our goal is to precisely model the non-linear power spectrum in cosmologies that include both massive neutrinos and modifications to $\Lambda$CDM. To do this we combine the halo model reaction for beyond-$\Lambda$CDM cosmologies \citep{Cataneo:2018cic} with that for massive neutrinos \citep{Cataneo:2019fjp}. 

The non-linear power spectrum, $P_{\rm NL}$, according to these prescriptions is the product of two key quantities
\begin{equation}
    P_{\rm NL}(k,z) = \mathcal{R}(k,z)  P^{\rm pseudo}_{\rm NL} (k,z) \, , \label{eq:nlps}
\end{equation}
with $\mathcal{R}(k,z)$ being the halo model reaction and $P^{\rm pseudo}_{\rm NL}(k,z)$ the {\it non-linear pseudo power spectrum}. The pseudo power spectrum describes a cosmology where the non-linear physics are governed by the $\Lambda$CDM model but whose linear clustering at the target redshift is tuned to match that of the `real', modified cosmology. 

\subsection{The halo model reaction: $\mathcal{R}$} 

The halo model reaction $\mathcal{R}$ then provides the non-linear corrections to the pseudo power spectrum coming from a non-zero neutrino mass and modifications to dark energy or gravity.
At its core, the halo model reaction is a ratio of halo model quantities - the real cosmology halo model prediction to the pseudo halo model prediction. Note that the benefit of using the pseudo cosmology as a reference is because this ensures the mass functions in both real and pseudo cosmologies (which have the same linear clustering) are similar. This allows a smoother transition between 2- and 1-halo terms. This was one of the issues in the standard halo model prescriptions \citep{Cooray:2002dia,Cacciato:2008hm,Giocoli:2010dm}. 

The reaction \B{including the effects of massive neutrinos \citep{Cataneo:2019fjp}} is given by 
\begin{equation}
    \mathcal{R}(k)=\frac{\left(1-f_{\nu}\right)^{2} P_{\mathrm{HM}}^{(\mathrm{cb})}(k)+2 f_{\nu}\left(1-f_{\nu}\right) P_{\mathrm{HM}}^{(\mathrm{cb} \nu)}(k)+f_{\nu}^{2} P_{\mathrm{L}}^{(\nu)}(k)}{P_{\mathrm{L}}^{(\mathrm{m})}(k)+P_{\mathrm{1h}}^{\mathrm{pseudo}}(k)} \, ,
    \label{eq:reaction}
\end{equation}
with $\rm (m) \equiv (cb+\nu) $, cb standing for CDM plus baryons and $\nu$ standing for massive neutrinos. \B{Here we have included the effects of massive neutrinos at the linear level in the real cosmology (numerator of \autoref{eq:nlps}) through the weighted sum of the non-linear halo model cb spectrum and massive neutrino linear spectrum \citep{Agarwal:2010mt}.} The components of the reaction are given by 
\begin{equation}
    P_{\mathrm{HM}}^{(\mathrm{cb} \nu)}(k) \approx \sqrt{P_{\mathrm{HM}}^{(\mathrm{cb})}(k) P_{\mathrm{L}}^{(\nu)}(k)} \, , 
\end{equation}
\begin{equation}
    P_{\mathrm{HM}}^{(\mathrm{cb})}(k) = \left[(1-\mathcal{E}) e^{-k / k_{\star}}+\mathcal{E}\right]  P_{\mathrm{L}}^{(\mathrm{cb})}(k)+P_{\mathrm{1h}}^{(\mathrm{cb})}(k) \, . \label{eq:1hcb} 
\end{equation}
Here we have added in the scale $k_{\star}$ and boost/suppression factor $\mathcal{E}$ \B{first introduced in C19} that have been shown to help the transition between 1- and 2-halo regimes in modified gravity theories. The linear spectra for CDM and baryons ($P_{\mathrm{L}}^{(\mathrm{cb})}$), massive neutrinos ($P_{\mathrm{L}}^{(\nu)}$) and total matter ($P_{\mathrm{L}}^{(\mathrm{m})}$) are provided by {\tt MGCAMB} \citep{Zucca:2019xhg,2011JCAP...08..005H,2009PhRvD..79h3513Z} for a particular modified gravity model including massive neutrinos or by {\tt CAMB} \citep{Lewis:2002ah} for $w \nu$CDM cosmologies. The 1-halo terms \B{for the real and pseudo cosmologies} are 
\begin{equation}
    P_{1 \mathrm{h}}^{\mathrm{(cb)}}(k)=\int \mathrm{d} \ln M \, n_{\mathrm{cb}}^{\rm MG}(M)\left(\frac{M}{\bar{\rho}_{\mathrm{cb}}}\right)^{2}|u_{\mathrm{cb}}^{\rm MG}(k, M)|^{2} \, , 
\end{equation}
\begin{equation}
    P_{1 \mathrm{h}}^{\mathrm{pseudo}}(k)=\int \mathrm{d} \ln M \, n^{\mathrm{pseudo}}(M)\left(\frac{M}{\bar{\rho}_{\mathrm{m}}}\right)^{2}|u^{\mathrm{pseudo}}(k, M)|^{2} \, , 
\end{equation}
where $\bar{\rho}$ is the background density for the relevant matter species and $u(k,M)$ is the Fourier transform of the halo density profile. The halo mass functions are given by
\begin{equation}
    n_{\mathrm{cb}}^{\rm MG}(M) = \frac{\bar{\rho}_{\mathrm{cb}}}{M}[\nu' f(\nu')] \frac{\mathrm{d} \ln \nu'}{\mathrm{d} \ln M} \, ,
\end{equation}
\begin{equation}
    n^{\mathrm{pseudo}}(M) = \frac{\bar{\rho}_{\mathrm{m}}}{M}[\nu'' f(\nu'')] \frac{\mathrm{d} \ln \nu''}{\mathrm{d} \ln M} \, .
\end{equation}
The peak-heights are  defined as $\nu' = \delta_{\mathrm{sc, cb}}^{\rm MG}(M)/\sigma_{\mathrm{cb}}^{\Lambda}(R_{\mathrm{cb}}(M))$ and $\nu'' = \delta_{\mathrm{sc,m}}^{\Lambda}/\sigma_{\mathrm{m}}^{\rm MG}(R_{\mathrm{m}}(M))$.  $\delta_{\mathrm{sc,cb}}^{\rm MG}(M)$ is obtained from solving the modified spherical collapse equations in the thin-shell approximation using only CDM and baryons to source the gravitational potential. On the other hand, $\delta_{\mathrm{sc}}^{\Lambda}$ is obtained by solving the standard, $\Lambda$CDM spherical collapse equations using the total matter density to source the gravitational potential. {\chh These definitions can be understood as follows. For the real cosmology, $\nu'$, we follow \cite{Hagstotz:2019}. The  MG effects are encoded in the initial collapse density through the spherical collapse computation. Here we linearly extrapolate the initial over-density using a $\Lambda$CDM growth following \cite{Cataneo:2018cic}. To ensure no evolutionary dependence on $\Lambda$CDM quantities we must then use $\sigma^{\Lambda}$ to preserve the initial peak statistic. Note that we assume at early times the linear spectrum in MG and $\Lambda$CDM are equivalent. Secondly, the pseudo peak-height follows from the definition of such a cosmology: a massless-neutrino $\Lambda$CDM cosmology with the linear power spectrum provided by the total matter power spectrum of the MG cosmology \citep[see][for details]{Cataneo:2018cic,Cataneo:2019fjp}. This simply means that the linear MG spectrum must be used for the linear mass variance $\sigma^{\rm MG}$, while the (non-linear) spherical collapse uses $\Lambda$CDM physics. Lastly, we use the {\it cold dark matter prescription} first introduced by \cite{Costanzi:2013} and later applied to $f(R)$ gravity cosmologies by \cite{Hagstotz:2019} to account for the effect of massive neutrinos on the halo number density in the real cosmology.}

The halo mass $M$ can be related to the radius $R$ by 
\begin{equation}
    M_{i}(R) = \frac{4\pi}{3} R_i^3 \bar{\rho}_{i,0} \, , 
\end{equation}
where the index $i\in \{cb,m\}$. For the variance of the mass fluctuations one has 
\begin{equation}
    [\sigma_{\mathrm{cb}}^{\Lambda}(R)]^2 = \int \frac{\mathrm{d}^{3} k}{(2 \pi)^{3}}|\tilde{W}(k R)|^{2} P_{\mathrm{L}}^{\rm (cb),\Lambda}(k),
\end{equation}
\begin{equation}
    [\sigma_{\mathrm{m}}^{\rm MG}(R)]^2 = \int \frac{\mathrm{d}^{3} k}{(2 \pi)^{3}}|\tilde{W}(k R)|^{2} P_{\mathrm{L}}^{\rm (m)}(k).
\end{equation}
$P_{\mathrm{L}}^{\rm (cb),\Lambda}(k)$ is the linear CDM plus baryon power spectrum in the real cosmology but without the modification to gravity, i.e. a $\Lambda$CDM cosmology with massive neutrinos.  

In this work we follow the procedure of C19: we use a Sheth-Torman mass function \citep{Sheth:1999mn,Sheth:2001dp}, a standard power law concentration-mass relation \citep[see for example][]{Bullock:1999he} and the halo density profile described in \cite{Navarro:1996gj}. We describe what inaccuracies these prescriptions incur in \autoref{sec:summary}. \B{Further, we also use the C19 definition for the virial radius. As in C19, the effect of modified gravity and/or massive neutrinos only enters these quantities through the collapse density and the virial theorem, determining the time of virialisation. The collapse density and linear variance of mass fluctuations also change the concentration-mass relation, and further, for $w$CDM cosmologies we introduce the factor motivated by \cite{Dolag:2003ui} in this relation, following C19. While the form of the mass function and density profile will be modified in non-standard cosmologies, keeping the standard forms serves as a good first approximation. We discuss going beyond this approximation in section.~\ref{sec:summary}.}

One can derive the $\mathcal{E}$ parameter in \autoref{eq:1hcb} as the limit
\begin{equation}
    \mathcal{E} = \frac{(1-f_\nu)^2 P_{1 \mathrm{h}}^{\mathrm{(cb)}}(k \rightarrow 0)}{P_{1 \mathrm{h}}^{\mathrm{pseudo}}(k \rightarrow 0)} 
    \, .
\end{equation}
To calculate $k_\star$ that appears in \autoref{eq:1hcb}, we need to use the 1-loop standard perturbation theory (SPT) prediction for the reaction. In particular we must solve $\mathcal{R}(k_0) = \mathcal{R}_{\rm SPT}(k_0)$. We do this at the wavenumber $k_0 = 0.06 \, h/\rm Mpc$ which is small enough to ensure the validity of the 1-loop SPT predictions, following C19. The SPT reaction is given by
\begin{align}
\label{eq:sptreaction}
    &\mathcal{R}_{\rm SPT}(k_0) = \\ \nonumber
    &\frac{(1-f_\nu)^2 P^{\rm (cb)}_0  + 2f_\nu(1-f_\nu)\sqrt{ P^{\rm (cb)}_0 P_{\rm L}^{(\nu)}(k_0)} + f_\nu^2 P_{\rm L}^{(\nu)}(k_0)}{P^{\rm pseudo}_0 } \, ,
\end{align}
with 
\begin{align}
P^{\rm (cb)}_0 &= P_{\rm SPT}^{\rm (cb)}(k_0) + P_{\rm 1h}^{\rm (cb)}(k_0) \nonumber \, , \\
P^{\rm pseudo}_0  &= P_{\rm SPT}^{\rm pseudo}(k_0) + P_{\rm 1h}^{\rm pseudo}(k_0) \, ,
\end{align}
where $P_{\rm SPT}^{\rm (cb)}(k_0)$ is computed following \cite{Saito:2009ah} with $P_{\mathrm{L}}^{(\mathrm{cb})}(k)$ for the MG + $M_\nu$ cosmology. $M_\nu$ is the sum of the neutrino masses, $M_\nu \equiv \sum m_\nu$, where $m_\nu$ is the mass of the individual species. The 1-loop spectrum is given by
\begin{equation} \label{eq:SPT}
    P_{\rm SPT}^{\rm (cb)}(k) = P_{\mathrm{L}}^{(\mathrm{cb})}(k) + P_{22}^{(\mathrm{cb})}(k) + P_{13}^{(\mathrm{cb})}(k) \, .
\end{equation}
Including both MG and massive neutrinos in the 1-loop computations for \autoref{eq:SPT} can be done as in \cite{Wright:2019qhf} as follows 
\begin{align}
P^{\rm (cb)}_{22}(k) =& 2 \frac{k^3}{(2\pi)^2}\int_0^{\infty}r^2 \mathrm{d}r \int_{-1}^1  \nonumber \\ 
& \times P_{{\rm L}}^{\rm (cb)}(kr)P_{{\rm L}}^{\rm (cb)}(k\sqrt{1+r^2-2rx})
\nonumber \\ 
&\times \frac{F_{2, {\rm MG}}^2(k, r, x)}{F_{1, {\rm MG}}^2(kr)F_{1, {\rm MG}}^2(k\sqrt{1+r^2-2rx})} \mathrm{d}x~,
 \label{eq:loop22}  \\
P^{\rm (cb)}_{13}(k) =& 6 \frac{k^3}{(2\pi)^2} P_{{\rm L}}^{\rm (cb)}(k) \int_0^{\infty} 
\nonumber \\ 
& \times r^2 P_{{\rm L}}^{\rm (cb)}(kr) \frac{F_{3, {\rm MG}}(k, r, x)}{F_{1, {\rm MG}}(k)F_{1, {\rm MG}}^2(kr)} \mathrm{d}r~,
\label{eq:loop13}
\end{align}
where again, $P_{{\rm L}}^{\rm (cb)}$ is taken from {\tt MGCAMB} and $F_i$, $i\in\{1,2,3\}$, being the 1st, 2nd and 3rd order SPT over-density kernels. No massive neutrino effects are included in the modified SPT kernels $F_i$ and they are computed using only CDM and baryons as sources to the gravitational potential. The kernels are computed as described in \cite{Bose:2016qun}. This massless-neutrino approximation for the SPT kernels was validated against simulations in the SPT regime of validity in \cite{Wright:2019qhf}.

The SPT pseudo computation is given by 
\begin{align}
    P_{\rm SPT}^{\rm pseudo}(k) = P_{\rm L}^{\rm (m)} (k) + P_{22}^{\rm pseudo}(k) + P_{13}^{\rm pseudo}(k) \, , 
\end{align}
where we {\it do not} use the `no-screening' approximation as in C19. Here the 22 and 13 loop terms are calculated as in \autoref{eq:loop22} and \autoref{eq:loop13} with all cb spectra replaced by the total matter spectra in the real, MG + $M_\nu$ cosmology and solving the 1st, 2nd and 3rd order SPT kernels without a modification to the Poisson equation, i.e. we replace $F_{i,\rm MG} \rightarrow F_{i,\rm \Lambda CDM}$ and $P^{\rm (cb)}_L \rightarrow P_L^{\rm (m)}$ in \autoref{eq:loop22} and \autoref{eq:loop13}.

We note that the formulation outlined in \autoref{eq:reaction} has  the property that in the limit $f_\nu \rightarrow 0$ we recover the results of C19, while in the case of no modification to gravity one gets $\mathcal{R}(k_0) \approx 1$ as expected from \cite{Cataneo:2019fjp}, meaning we do not need SPT for $\nu\Lambda \rm CDM$ nor $\nu w$CDM cosmologies. \B{Indeed, these parameters (and the modification of the real 2-halo term), were introduced to account for new mode-couplings and screening mechanisms when moving from linear to non-linear power spectrum. While these parameters naturally go to their GR values for the $\nu w$CDM cosmologies we have considered, the code includes a flag for the inclusion or exclusion of modified gravity (and of SPT) to ensure theoretical consistency and protect against numerically-related deviations of $k_\star$ and $\mathcal{E}$ in cosmologies without modifications to gravity.}

\subsection{Theoretical accuracy and the pseudo power spectrum } 

We note that the accuracy of \autoref{eq:reaction} relies on the accuracy of both $P_{\rm NL}^{\rm pseudo}$ and $\mathcal{R}$. It was shown in C19 that $\mathcal{R}$ is accurate at the $1\%$ level at $k\leq 1h/{\rm Mpc}$ for all considered beyond-$\Lambda$CDM cosmologies. In \cite{Cataneo:2019fjp}, it was shown that $\mathcal{R}$ is accurate at the $1\%$ level at $k\leq 10h/{\rm Mpc}$ for $\nu \Lambda$CDM cosmologies with $M_{\nu}\leq0.4$eV. These estimates all made use of an $N$-body simulated $P_{\rm NL}^{\rm pseudo}$ which introduces negligible inaccuracy to the final $P_{\rm NL}$ prediction. We do not have the benefit of such accurate pseudo spectra for the cosmologies we are considering in this work.

Instead, we model the pseudo spectrum using the halo model inspired fitting formula of \cite{Mead:2020vgs} which is accurate at the $5\%$-level for $k<10h/{\rm Mpc}$ which sets the accuracy for our theoretical predictions for $k\leq 1h/{\rm Mpc}$. Above this we also incur inaccuracies from $\mathcal{R}$, largely attributed to inaccuracies in the halo mass function and concentration-mass relation of the real and pseudo cosmologies. 

Our adopted prescription for the pseudo spectrum can be computed using the publicly available {\tt HMCode2020} \footnote{Download {\tt HMCode2020}: \url{https://github.com/alexander-mead/HMcode}}. In practice this is achieved by giving the modified linear (total matter) spectrum to {\tt HMCode2020} with all parameters set to their $\Lambda$CDM values. 

\B{In the ideal case, one would make use of a bespoke emulator as suggested in \cite{Giblin:2019iit}. Such an emulator would be able to match a large range of target linear spectra. For models which only introduce a scale-independent growth modification at the linear level, one could conceivably use standard emulators based on GR such as Bacco or  EuclidEmulator2 and adjust the linear spectrum amplitude through $\sigma_8$ or $A_s$ parameters, to match the target linear amplitude, but for scale-dependent theories, and indeed for the inclusion of massive neutrinos, matching the target linear spectrum becomes non-trivial for these emulators. This is both because of the non-trivial shape of the linear power spectrum for some beyond-$\Lambda$CDM theories such as $f(R)$ and any theory combined with massive neutrinos, but also because of the restricted range in parameter space of the GR emulators.}

\B{One general issue of the emulator approach, even with a bespoke pseudo emulator, is that of interpolation error. A finite number of nodes will result in some inaccuracy for cosmologies between the nodes. In \cite{Giblin:2019iit} they find that a few hundred nodes are sufficient for $\sim 2$\% accuracy down to $k=10h/{\rm Mpc}$ for $z\leq 1$. This can be further ameliorated by sensibly reducing the emulated parameter volume based on the posterior distribution of a particular statistic of interest \citep[see, e.g.,][for possible strategies]{DeRose:2019,Rogers:2019}.}

\B{On the other hand, the non-linear pseudo power spectrum as modelled by a fitting function such as {\tt HMCode2020} doesn't have parameter range issues nor interpolation errors as it uses the exact target linear spectrum as input. Despite this, it does suffer from a lack of precision coming from its relatively low number of free parameters which are also fit to a relatively limited set of $N$-body simulations.}

\B{In the ideal case, a sufficiently accurate and comprehensive emulator as proposed in \cite{Giblin:2019iit} would be used. We comment more on the ideal setup in \autoref{sec:summary}.}

In the next section we test the combination of these effects against $N$-body simulations and state-of-the-art emulators. 


\section{Accuracy validation of the modelling against simulations and emulators}\label{sec:results} 
We compare the theoretical prediction given by \autoref{eq:nlps} to highly accurate estimates for the non-linear matter power spectrum. Specifically we consider sets of simulation measurements and state-of-the-art emulators for beyond-$\Lambda$CDM cosmologies. 

\subsection{Massive neutrinos in beyond-$\Lambda$CDM}\label{sec:neutrinos}

We will consider two beyond-$\Lambda$CDM scenarios with the inclusion of massive neutrinos:
\begin{enumerate}
    \item 
    The Hu-Sawicki \citep{Hu:2007nk} $f(R)$ gravity model, which induces scale-dependent growth and comes with an environment-dependent screening mechanism, allowing the recovery of GR within the solar system. $|f_{\rm R0}|$ is the free parameter of the theory and is the value of the scalar field today. 
    \item
    An evolving dark energy parametrised as in \cite{Chevallier:2000qy} and \cite{Linder:2002et}, which we will denote as $w$CDM. This comes with two free parameters: the equation of state of dark energy today ($w_0$) and one governing its time evolution as $w(a) = w_0 + (1-a)w_a$.
\end{enumerate}

\subsubsection{$f(R)$ gravity with massive neutrinos}\label{sec:resultsfr}
For this scenario we will compare the predictions of \autoref{eq:nlps} 
with the DUSTGRAIN-\textit{pathfinder} simulations \citep{Giocoli:2018gqh}.
The DUSTGRAIN-\textit{pathfinder} simulations are part of a suite of cosmological runs designed to sample a variety of combinations of modified gravity and massive neutrinos cosmologies.
The runs have been performed with the \texttt{MG-GADGET} code \citep{puchwein13}, and subsequently post-processed for different studies \citep{girelli20,corasaniti20} including weak lensing light-cones using the \texttt{MapSim} routine \citep{giocoli15,hilbert20}. Weak lensing observables have been validated and studied in a variety
of works going from standard two point statistics and PDF \citep{boyle20} to more complex machine learning 
analyses \citep{merten19,peel19}. In addition, the runs have been used to study halo clustering \citep{garciaf19} and void properties \citep{contarini21}, 
respectively.  

These simulations have a baseline cosmology of $h = 0.6731$, $n_s = 0.9658$,  $\Omega_m = 0.31345$, $\Omega_b = 0.0491$ and $A_s = 2.2 \times 10^{-9}$; following the evolution of $768^3$ dark 
matter particles -- doubled in presence of massive neutrinos -- in a volume of 
750 Mpc$/h$ by side with periodic boundary conditions. Initial conditions have been generated at $z=99$ from a random realization 
of an initial power spectrum computed using \texttt{CAMB}.

\begin{figure}
\centering
  \includegraphics[width=0.48\textwidth,height=0.25\textwidth]{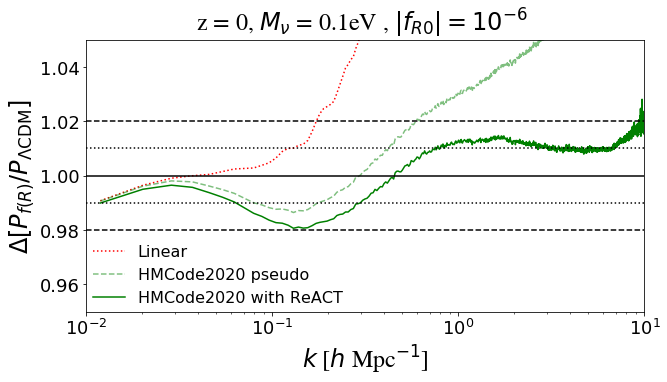}
    \includegraphics[width=0.48\textwidth,height=0.25\textwidth]{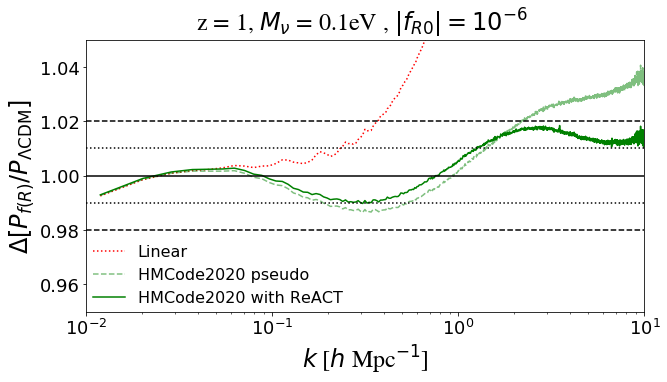}
    \\
  \caption[CONVERGENCE]{Comparison of theoretical predictions to DUSTGRAIN-\textit{pathfinder} measurements in $f(R)$ + $M_\nu$ (case (a) in main text) with $|f_{\rm R0}|=10^{-6}$ and $M_\nu=0.1$eV.  We compare the ratio of the $f(R)$ + $M_\nu$ $P(k)$ to the $\Lambda$CDM $P(k)$, in the two cases. Top is $z=0$ and bottom is $z=1$. We show linear (red dotted), {\tt HMCode2020} pseudo (green dashed) and {\tt HMCode2020} pseudo with reaction (green solid) predictions.}
\label{mnu01f6}
\end{figure}

\begin{figure}
\centering
  \includegraphics[width=0.48\textwidth,height=0.25\textwidth]{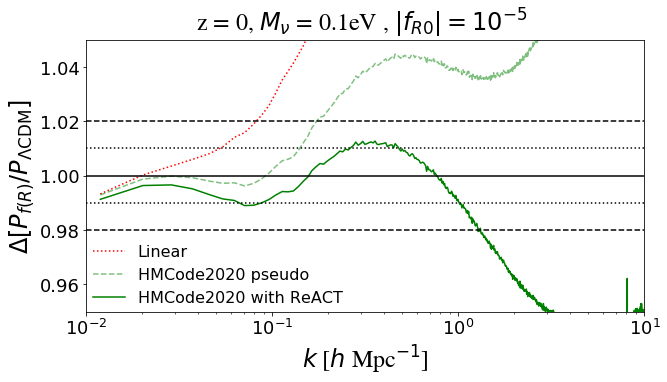}
    \includegraphics[width=0.48\textwidth,height=0.25\textwidth]{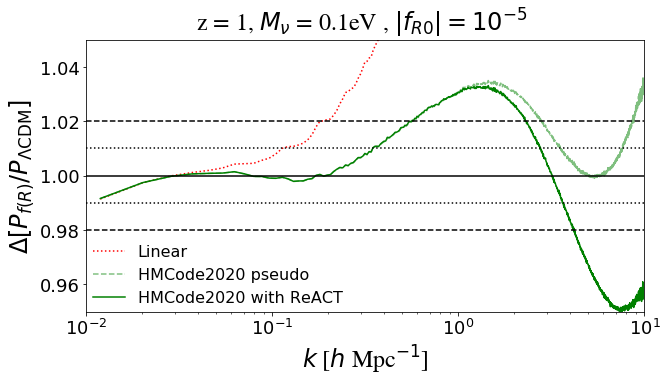}
    \\
  \caption[CONVERGENCE]{Comparison of theoretical predictions to DUSTGRAIN-\textit{pathfinder} measurements in $f(R)$ + $M_\nu$ (case (b) in main text) with $|f_{\rm R0}|=10^{-5}$ and $M_\nu=0.1$eV.  We compare the ratio of the $f(R)$ + $M_\nu$ $P(k)$ to the $\Lambda$CDM $P(k)$, in the two cases. Top is $z=0$ and bottom is $z=1$. We show linear (red dotted), {\tt HMCode2020} pseudo (green dashed) and {\tt HMCode2020} pseudo with reaction (green solid) predictions.}
\label{mnu01f5}
\end{figure}

\begin{figure}
\centering
  \includegraphics[width=0.48\textwidth,height=0.25\textwidth]{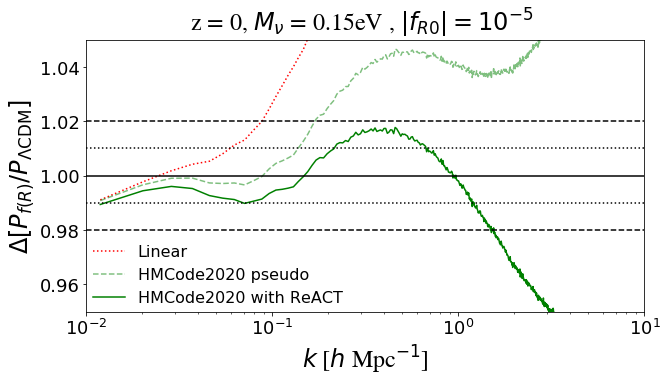}
    \includegraphics[width=0.48\textwidth,height=0.25\textwidth]{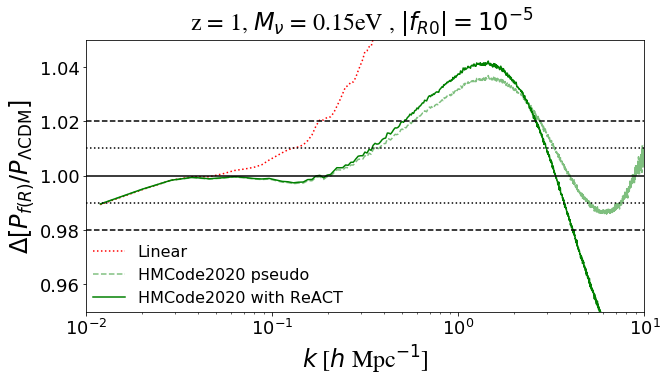}
    \\
  \caption[CONVERGENCE]{Comparison of theoretical predictions to DUSTGRAIN-\textit{pathfinder} measurements in $f(R)$ + $M_\nu$ (case (c) in main text) with $|f_{\rm R0}|=10^{-5}$ and $M_\nu=0.15$eV.  We compare the ratio of the $f(R)$ + $M_\nu$ $P(k)$ to the $\Lambda$CDM $P(k)$, in the two cases. Top is $z=0$ and bottom is $z=1$. We show linear (red dotted), {\tt HMCode2020} pseudo (green dashed) and {\tt HMCode2020} pseudo with reaction (green solid) predictions. }
\label{mnu015f5}
\end{figure}

In this work, we consider three models in the 2D parameter space $\{f_{\rm R0}$,$M_\nu \}$ of increasing  
deviation from  $\Lambda$CDM:

\begin{enumerate}
    \item[(a)] 
    Low:  $M_\nu = 0.1$eV ($\Omega_\nu = 0.00238$, $\Omega_{\rm cdm} = 0.26197 $), $|f_{\rm R0}| = 10^{-6}$. 
    \item[(b)] 
    Medium: $M_\nu = 0.1$eV ($\Omega_\nu = 0.00238$, $\Omega_{\rm cdm} = 0.26197 $), $|f_{\rm R0}| = 10^{-5}$. 
    \item[(c)] 
    High: $M_\nu = 0.15$eV ($\Omega_\nu = 0.00358$, $\Omega_{\rm cdm} = 0.26077 $), $|f_{\rm R0}| = 10^{-5}$. 
\end{enumerate}
In \autoref{mnu01f6}, \autoref{mnu01f5} and \autoref{mnu015f5} we show the relative change of the matter power spectrum in the modified cosmologies to $\Lambda$CDM, i.e. the ratio of the ratio $P_{\rm f(R)}(k)/P_{\Lambda{\rm CDM}}$ between the theoretical prediction and the simulation measurement, for cases (a), (b) and (c) respectively. For reference we also show the linear theory prediction and the prediction without the reaction, i.e. $ P^{\rm pseudo}_{\rm NL} (k,z)$. 

We find that for all cases the halo model reaction with the pseudo spectrum prescription of \cite{Mead:2020vgs} is $2(4)\%$ accurate for all cases at scales $k\leq 1(2)h/{\rm Mpc}$ at $z=0$. Considering highly non-linear scales, the low deviation case, (a), is $2\%$ accurate at scales $k\leq 10h/{\rm Mpc}$ for $z=0$ and $z=1$. At $z=0(1)$, case (b) and (c) show up to 5(3)\% and 6(4)\% deviations respectively within $k\leq 5h\ {\rm Mpc}$. Note that this is the limiting accuracy of {\tt HMCode2020}, but without a measurement of the true pseudo spectrum from simulations we cannot discriminate between inaccuracies in the reaction, $\mathcal{R}$, and the pseudo spectrum, $ P^{\rm pseudo}_{\rm NL}$.

To investigate this issue, we have run a set of COmoving Lagrangian Acceleration (COLA) simulations, including a set for the pseudo cosmology, \B{using the approach from \cite{Winther:2017jof} and \cite{Wright:2017dkw} that is implemented in the COLA code {\tt FML}\footnote{Download {\tt FML}: \url{https://github.com/HAWinther/FML/tree/master/FML/COLASolver}}}. These results are presented in appendix~\ref{app:cola}. \B{For case (b) we} find that the accuracy of the pseudo-COLA spectrum application is less than $~2\%$ for $k\leq 3h/{\rm Mpc}$ at $z=1$, \B{ while for (a) it is less than $~3\%$}. \B{This indicates that} the bump shown by the solid green lines in the bottom panels of \autoref{mnu01f5} and \autoref{mnu015f5} \B{partially} comes from {\tt HMCode2020}. \B{We comment on this further in appendix~\ref{app:cola} and} await full $N$-body simulations for the pseudo cosmology to \B{investigate this issue further.}

Finally, we note that the tilt (and non-unity) observed in the comparisons at large scales ($k\leq 0.02h/{\rm Mpc}$) is likely due to relativistic effects for massive neutrino cosmologies, included in the {\tt MGCAMB} predictions but not in the simulations. Similar effects were observed in \cite{Tram:2018znz,Massara:2014kba}. We also note a similar trend when comparing to the BAHAMAS simulations in \autoref{sec:baryons}. 

\subsubsection{$w$CDM with massive neutrinos}

For this scenario we will compare the predictions of \autoref{eq:nlps}  with predictions given by the Bacco emulator\footnote{Download Bacco: \url{http://www.dipc.org/bacco/emulator.html}} \citep{2020arXiv200406245A} and the EuclidEmulator2\footnote{Download EuclidEmulator2: \url{https://github.com/miknab/EuclidEmulator2}} \citep{Knabenhans:2020gdo}. The Bacco  (EuclidEmulator2) is expected to be accurate at the $3 (\leq 1)\%$ level down to $k=5 (10) h$/Mpc.

We adopt the base cosmology $h = 0.7$, $n_s = 0.972$,  $\Omega_m = 0.2793$ and $\Omega_b = 0.0463$. \B{We then take 6 samples from the overlapping parameter space of both emulators in $\{M_\nu, w_0,w_a\} \in \{ [0,0.15]eV , [-1.15,-0.85], [-0.3,0.3]\}$. For these 6 cosmologies, we compare the EuclidEmulator2 emulator predictions with the Bacco emulator, the Halofit fitting formula \citep{Takahashi:2012em},  the stand-alone {\tt HMCode2020} predictions and finally with \autoref{eq:nlps} (a non-linear pseudo spectrum given by {\tt HMCode2020} combined with the halo model reaction given in \autoref{eq:reaction}).}

\begin{figure*}
\centering
  \includegraphics[width=\textwidth,height=0.5\textwidth]{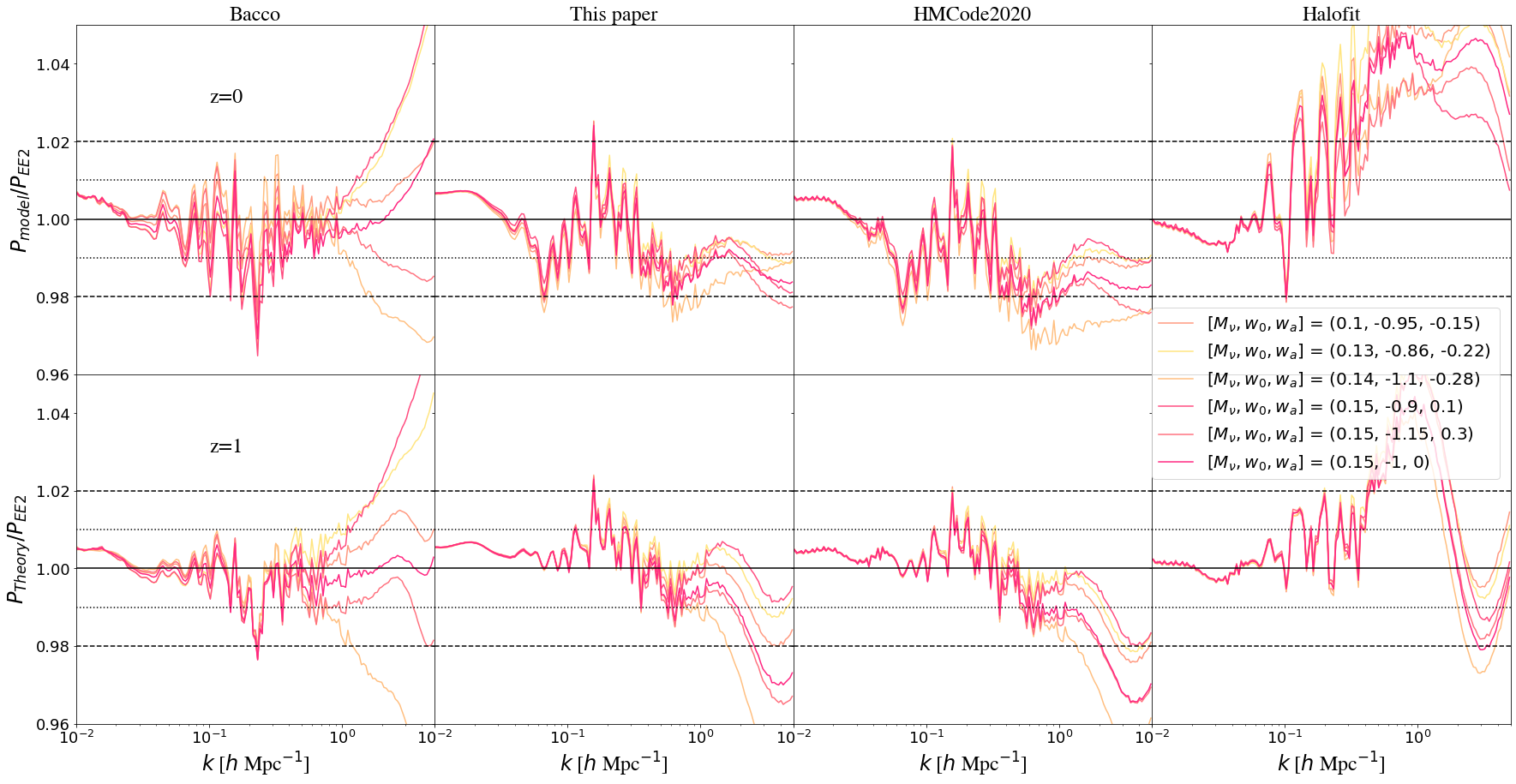}
  \caption[CONVERGENCE]{Each column shows the ratio of a particular prescription for the non-linear matter power spectrum to the EuclidEmulator2 prediction. The various lines represent cosmologies with a different choice of sum of neutrino mass $M_{\rm \nu}$ and dark energy equation of state parameters $\{w_0,w_a\}$. Top panels show the results for $z=0$ and bottom panels for $z=1$.}
\label{fig:wcdmall}
\end{figure*}

\B{We show the ratios of the various prescriptions for $P_{\rm NL}(k)$ to the EuclidEmulator2 in \autoref{fig:wcdmall}. We find that an {\tt HMCode2020} prescription for the pseudo spectrum combined with the halo model reaction offers 1(2)\% consistency with  EuclidEmulator2 at $z=1(0)$ for $k\leq1h/{\rm Mpc}$ for the full range of $\nu w$CDM cosmologies considered. The accuracy remains at the $2\%$ level down to $k=5h/{\rm Mpc}$ at z=0 for all cosmologies but worsens to the $5\%$ level for large values of $|w_a|$ at $z=1$. }

\B{The accuracy of our approach is comparable to {\tt HMCode2020}'s $\nu w$CDM predictions over the full range of cosmologies while Bacco deviates largely for large modifications to a constant dark energy which is consistent with the results shown in Appendix A of \cite{Contreras:2020kbv}. Halofit on the other hand shows up to 5\% disagreement for all cosmologies beyond $k\approx 0.5h/{\rm Mpc}$.}

\begin{figure*}
\centering
  \includegraphics[width=\textwidth,height=0.5\textwidth]{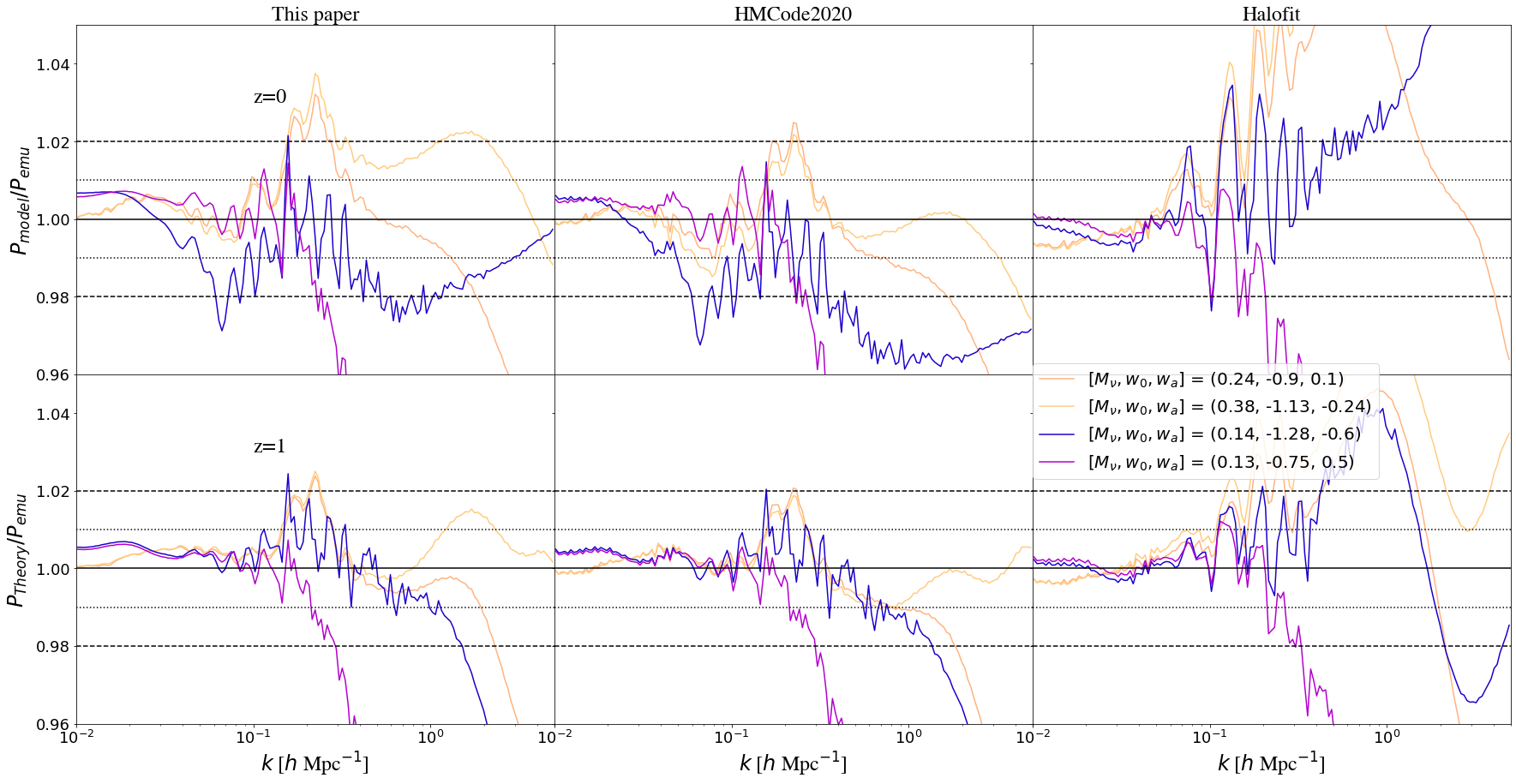}
  \caption[CONVERGENCE]{{\bf `Extreme' cases:} Each column shows the ratio of a particular prescription for the non-linear matter power spectrum to an emulator prediction. In particular, the orange curves are compared to the Bacco emulator while the purple curves to the EuclidEmulator2. The various lines represent cosmologies with a different choice of sum of neutrino mass $M_{\rm \nu}$ and dark energy equation of state parameters $\{w_0,w_a\}$. Top panels show the results for $z=0$ and bottom panels for $z=1$.}
\label{fig:ext}
\end{figure*}

\B{To further test the limitations of our modelling prescription, we have also investigated some extreme parameter choices which are well beyond the current cosmological constraints (see for example \cite{Aghanim:2018eyx}). These are shown in \autoref{fig:ext}. We consider two large neutrino mass cases which are permitted by the Bacco emulator range and two large deviations from a constant dark energy permitted by the Euclid emulator range. For most cases we find \autoref{eq:nlps} remains within $2\%$ of the emulators down to $k=1h/{\rm Mpc}$ at $z=0$ and $z=1$ with the exception of a case with $w_a=0.5$ which is at the edge of the range at which the current version of {\tt ReACT} is numerically stable ($w_a\leq 0.8$). This case deviates by more than $2\%$ from  EuclidEmulator2 at $k\approx 0.3h/{\rm Mpc}$. Again {\tt HMCode2020} shows similar consistency with the emulators in this range overall while halofit disagrees with the emulators beyond 5\% above $k\approx 5h/{\rm Mpc}$. }

\B{In all cases considered we note that the reaction adds significant accuracy to the {\tt HMCode2020} pseudo spectrum, especially above $k=1h$/Mpc.}

\subsection{Including baryonic feedback}\label{sec:baryons}

We now look to include baryonic effects in our predictions. To do this we make use of the feedback modelling included in {\tt HMCode2020}. We compare to the BAHAMAS suite of cosmological hydrodynamical simulations \citep{McCarthy:2016mry,McCarthy:2017csu}. In particular we will make use of the $\nu\Lambda$CDM and $\nu w$CDM BAHAMAS simulations described in \cite{vanDaalen:2019pst} and \cite{Pfeifer:2020jct} respectively. The {\tt HMCode2020} feedback parameter is set to $\log_{10}(T_{\rm AGN}/K) = 7.8$ which was fit to the $\nu\Lambda$CDM BAHAMAS simulations \citep{Mead:2020vgs}.

The effects of feedback from AGN and stellar sources (supernovae) are incorporated in the BAHAMAS simulations using subgrid models that have a number of free parameters.  As described in \citet{McCarthy:2016mry}, these parameters were adjusted so that the simulations reproduce the local galaxy stellar mass function and the hot gas mass fractions of galaxy groups and clusters, thus ensuring that massive haloes (which contribute the most to the matter power spectrum) have the correct baryon fractions.  Note that \citet{vanDaalen:2019pst} have shown that having the correct baryon fractions is key to obtaining a realistic impact of baryons on the total matter power spectrum.  Furthermore, both \citet{McCarthy:2017csu} and \citet{vanDaalen:2019pst} have shown that the impact of baryons on the power spectrum is only very weakly dependent on cosmology (primarily through the universal baryon fraction, $\Omega_b/\Omega_m$), such that recalibration of the feedback is generally unnecessary when varying cosmology.  Indeed, for the  $\nu\Lambda$CDM and $\nu w$CDM simulations we use here, the feedback prescription was left unchanged from the fiducial BAHAMAS run but it was verified that the relative impact of the power spectrum (or the predicted baryon fractions) did not change by more than about a percent.  Finally, we note that \citet{vanDaalen:2019pst} have demonstrated that on very large scales (where the impact of baryons is expected to be unimportant) the ratio of the hydro simulations to their dissipationless counterparts converges to typically better than 0.1\% accuracy, which might be viewed as the numerical accuracy of the predicted (relative) power spectra.

\subsubsection{$\nu\Lambda$CDM} 
\label{sec:nulcdmbar}
\begin{figure}
\centering
  \includegraphics[width=0.48\textwidth,height=0.25\textwidth]{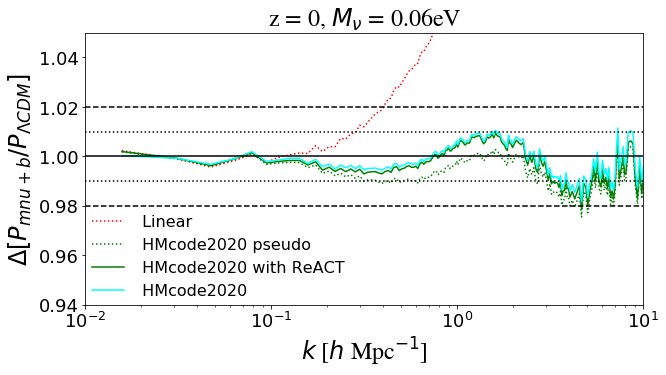}
    \includegraphics[width=0.48\textwidth,height=0.25\textwidth]{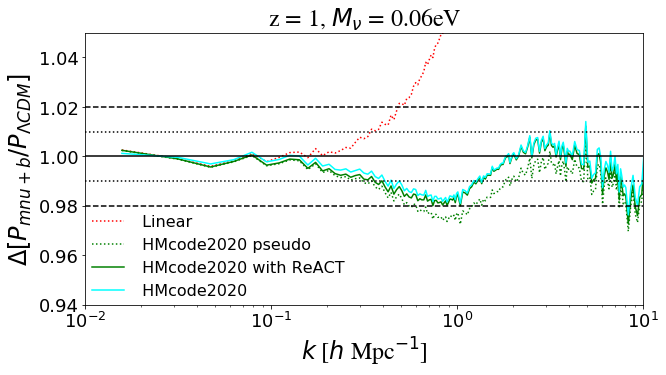}
    \\
  \caption[CONVERGENCE]{Comparison of theoretical predictions to BAHAMAS measurements in $\nu \Lambda$CDM (case (a) in main text) with $M_\nu = 0.06$eV.  We compare the ratio of the $\nu \Lambda$CDM $P(k)$ to the $\Lambda$CDM $P(k)$, in the two cases. Top is $z=0$ and bottom is $z=1$.  We show linear (red dotted), {\tt HMCode2020} pseudo (green dashed), {\tt HMCode2020} pseudo with reaction (green solid) and {\tt HMCode2020} (cyan solid) predictions.}
\label{baha06}
\end{figure}

\begin{figure}
\centering
  \includegraphics[width=0.48\textwidth,height=0.25\textwidth]{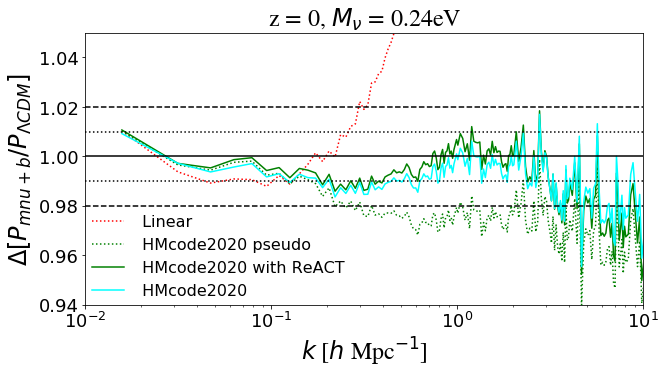}
    \includegraphics[width=0.48\textwidth,height=0.25\textwidth]{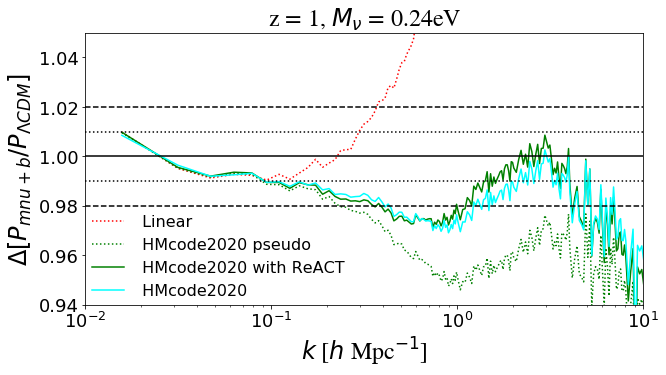}
    \\
  \caption[CONVERGENCE]{Comparison of theoretical predictions to BAHAMAS measurements in $\nu \Lambda$CDM (case (b) in main text) with $M_\nu = 0.24$eV.  We compare the ratio of the $\nu \Lambda$CDM $P(k)$ to the $\Lambda$CDM $P(k)$, in the two cases. Top is $z=0$ and bottom is $z=1$. We show linear (red dotted), {\tt HMCode2020} pseudo (green dashed), {\tt HMCode2020} pseudo with reaction (green solid) and {\tt HMCode2020} (cyan solid) predictions.}
\label{baha24}
\end{figure}

\begin{figure}
\centering
  \includegraphics[width=0.48\textwidth,height=0.25\textwidth]{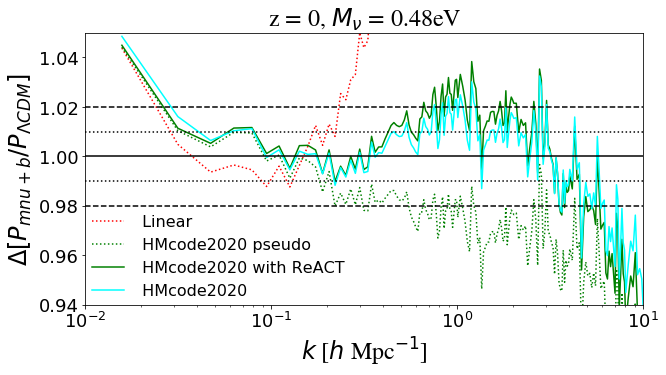}
    \includegraphics[width=0.48\textwidth,height=0.25\textwidth]{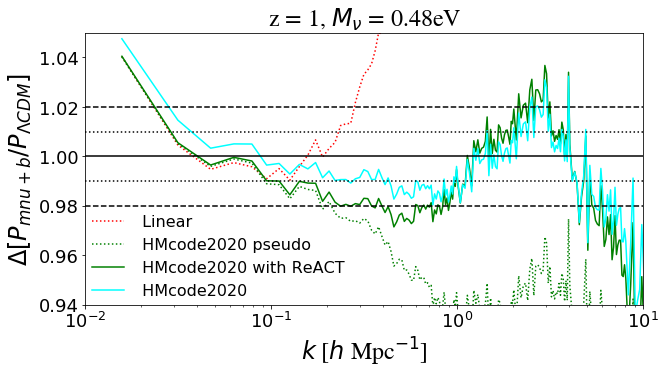}
    \\
  \caption[CONVERGENCE]{Comparison of theoretical predictions to BAHAMAS measurements in $\nu \Lambda$CDM (case (c) in main text) with $M_\nu = 0.48$eV.  We compare the ratio of the $\nu \Lambda$CDM $P(k)$ to the $\Lambda$CDM $P(k)$, in the two cases. Top is $z=0$ and bottom is $z=1$. We show linear (red dotted), {\tt HMCode2020} pseudo (green dashed), {\tt HMCode2020} pseudo with reaction (green solid) and {\tt HMCode2020} (cyan solid) predictions. }
\label{baha48}
\end{figure}

We consider the WMAP9 cosmology of this suite which adopts the baseline parameters $h = 0.7$, $n_s = 0.972$,  $\Omega_m = 0.2793$, $\Omega_b = 0.0463$ and $A_s = 2.392 \times 10^{-9}$. We consider 3 massive neutrino cases:
\begin{enumerate}
    \item[(a)] 
    Low mass: $M_\nu = 0.06$eV ($\Omega_\nu = 0.0013$, $\Omega_{\rm cdm} = 0.2317 $).
    \item[(b)]
    Medium mass: $M_\nu = 0.24$eV ($\Omega_\nu = 0.0053$, $\Omega_{\rm cdm} = 0.2277 $).
    \item[(c)]
    High mass: $M_\nu = 0.48$eV ($\Omega_\nu = 0.0105$, $\Omega_{\rm cdm} = 0.2225 $). 
\end{enumerate}
We again show the ratio of the quantity $P_{\rm mnu +b}(k)/P_{\Lambda{\rm CDM}}$, where `mnu+b' stands for the massive neutrino cosmology with baryonic effects, between the theoretical predictions and the simulation measurements\footnote{The $P_{\Lambda{\rm CDM}}$ quantity includes no baryonic \B{nor massive neutrino} effects  i.e. the simulation measurement is from a dark matter-only simulation as opposed to cases (a), (b) and (c) which are all made from hydrodynamical simulations with massive neutrinos.}. This is shown for cases (a), (b) and (c) in \autoref{baha06}, \autoref{baha24} and \autoref{baha48} respectively. For all cases we find that the halo model reaction combined with the {\tt HMCode2020} pseudo spectrum is $\leq 3\%$ accurate for $k\leq 5h/{\rm Mpc}$ for $z=0$ and $z=1$,  with the predictions generally being more accurate for lower neutrino mass and $z=0$. \B{Further, for all cases we find sub-percent agreement over all scales and redshifts between the stand-alone {\tt HMCode2020} and the halo model reaction with a pseudo {\tt HMCode2020} prediction.} We note that the feedback model of {\tt HMCode2020} is fit to the BAHAMAS simulations and so the high degree of accuracy is not surprising. 

\B{It is worth highlighting that the baryonic feedback model of {\tt HMCode2020} is implemented through another `reaction',  that of the dark matter spectrum to baryonic  effects. Thus, our predictions are applying two reactions independently: one for massive neutrinos and one for baryonic effects, both based on different conventions. Ideally, these would be combined consistently into a single reaction as we have done for massive neutrinos and modified gravity or dark energy. We leave this for future work. This being said, the results of \cite{Mead:2016zqy} show a high degree of independence between massive neutrino and baryonic effects down to $k=10h/{\rm Mpc}$ which is consistent with both the accuracy found in \autoref{baha06}, \autoref{baha24} and \autoref{baha48}, as well as the consistency between the single reaction application of {\tt HMCode2020} (cyan curves) and the double-reaction application {\tt HMCode2020}-pseudo together with the massive neutrino reaction.} 

Again, we also note the tilt (and non-unity ratio) observed in the comparisons at large scales are likely due to relativistic effects included in the linear power spectrum produced by {\tt MGCAMB}, as also noted in the DUSTGRAIN-\textit{pathfinder} simulation comparisons.

\subsubsection{$\nu w$CDM} 
We will make use of three simulations from this suite, all of which have $n_s =0.97$ and a neutrino mass of $M_\nu = 0.06$eV. The other cosmological parameters are detailed below:
\begin{enumerate}
    \item[(a)] 
    Non-phantom: $\Omega_m = 0.286$, $\Omega_b =0.0462$, $H_0=69.97$, $\sigma_8=0.819$ and $\{w_0,w_a\} = \{-0.67,-1.45\}$.
    \item[(b)]
    Phantom:  $\Omega_m = 0.309$, $\Omega_b =0.0501$, $H_0=67.25$, $\sigma_8=0.773$ and $\{w_0,w_a\} = \{-1.16,0.73\}$.
    \item[(c)]
     $\nu \Lambda$CDM: $\Omega_m = 0.294$, $\Omega_b =0.0476$, $H_0=68.98$ and $\sigma_8=0.802$ (dark matter only).
\end{enumerate}
In this subsection the theoretical predictions for cases (a) and (b) follow \autoref{eq:nlps} with $\mathcal{R}$ including both evolving dark energy and massive neutrino effects while $P^{\rm pseudo}_{\rm NL}$ again includes baryonic feedback effects. Case (c) will also follow \autoref{eq:nlps} but $\mathcal{R}$ will only model the effect of a non-zero neutrino mass and $P^{\rm pseudo}_{\rm NL}$ will include no baryonic feedback effects. \B{We again show the stand-alone {\tt HMCode2020} predictions too which supports $\nu w$CDM cosmologies.} 

We note that the simulation measurements in these cases are of the cold dark matter plus baryon (cb) power spectrum and do not include the massive neutrino contribution directly. The theoretical predictions have been adjusted accordingly. We have checked that for the ratios we plot, the theoretical predictions for the full matter spectrum and for the cb spectrum are almost identical.  

\begin{figure}
\centering
  \includegraphics[width=0.48\textwidth,height=0.25\textwidth]{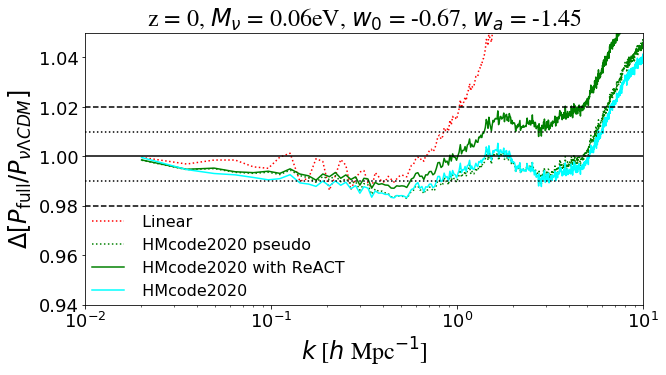}
    \includegraphics[width=0.48\textwidth,height=0.25\textwidth]{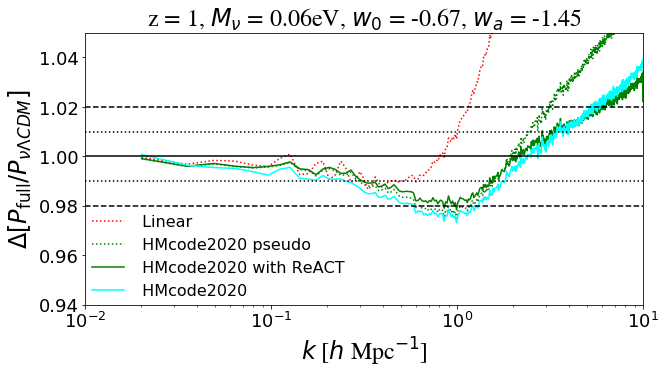}
    \\
  \caption[CONVERGENCE]{Comparison of theoretical predictions to BAHAMAS measurements in non-phantom $\nu w$CDM (case (a) in main text) with $M_\nu = 0.06$eV.  We compare the ratio of the $\nu w$CDM $P(k)$ to the $\nu \Lambda$CDM $P(k)$, in the two cases. Top is $z=0$ and bottom is $z=1$. We show linear (red dotted), {\tt HMCode2020} pseudo (green dashed), {\tt HMCode2020} pseudo with reaction (green solid) and {\tt HMCode2020} (cyan solid) predictions. }
\label{wcdmnu}
\end{figure}

\begin{figure}
\centering
  \includegraphics[width=0.48\textwidth,height=0.25\textwidth]{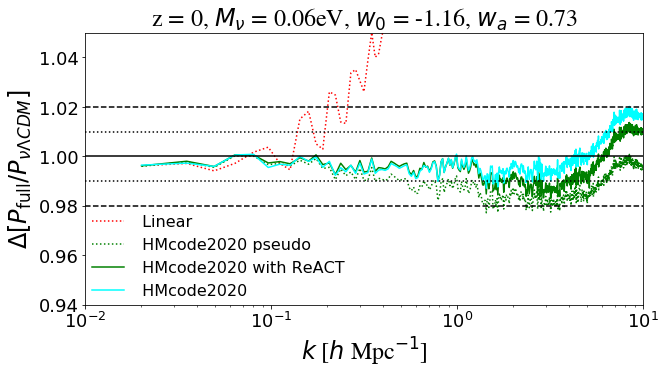}
    \includegraphics[width=0.48\textwidth,height=0.25\textwidth]{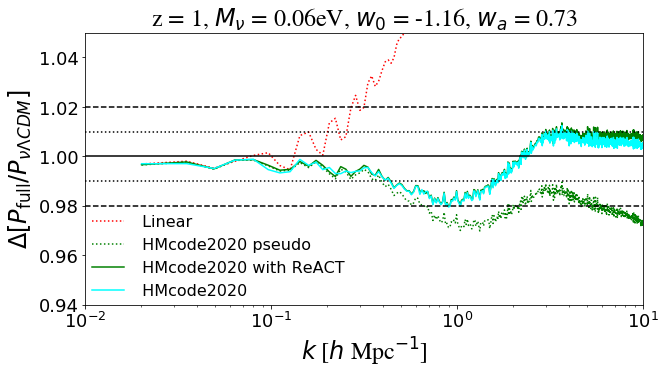}
    \\
  \caption[CONVERGENCE]{Comparison of theoretical predictions to BAHAMAS measurements in phantom $\nu w$CDM (case (b) in main text) with $M_\nu = 0.06$eV.  We compare the ratio of the $\nu w$CDM $P(k)$ to the $\nu \Lambda$CDM $P(k)$, in the two cases. Top is $z=0$ and bottom is $z=1$. We show linear (red dotted), {\tt HMCode2020} pseudo (green dashed), {\tt HMCode2020} pseudo with reaction (green solid) and {\tt HMCode2020} (cyan solid) predictions.}
\label{phantnu}
\end{figure}

The ratio of cases (a) and (b) to case (c) as given by theory to the same ratio as measured from the simulations are shown in \autoref{wcdmnu} and \autoref{phantnu} respectively. For both cases we find that the halo model reaction combined with a {\tt HMCode2020} pseudo is $\leq 2\%$ accurate for $k\leq 5h/{\rm Mpc}$ for $z=0$ and $z=1$. \B{As in the $\nu \Lambda$CDM cases shown in \autoref{sec:nulcdmbar}, the predictions of \autoref{eq:nlps} are generally in sub-percent agreement with the stand-alone {\tt HMCode2020} predictions.}

The lack of a $\Lambda$CDM simulation without massive neutrinos nor higher neutrino mass $\nu w$CDM simulations prevents us from investigating the accuracy of our combined massive neutrino and evolving dark energy theoretical prescription in more detail. We do however find that the accuracy demonstrated for these two cases is shown for all the $\nu w$CDM cosmologies outlined in \cite{Pfeifer:2020jct}, with the inclusion of the reaction $\mathcal{R}$ generally adding to the accuracy of \autoref{eq:nlps}. Further, this supports the reliability of modelling evolving dark energy and baryonic effects independently. 

\B{Finally, we note that the magnitude and shape of the reaction in our calculations inferred from \autoref{wcdmnu} is very similar to what was found in C19. In particular their DE5 case is very similar to our case (a), exhibiting a $2\%$ accuracy with their dark matter simulations up to $k\sim 5h/{\rm Mpc}$. This further indicates that the small neutrino mass, baryonic effects and variation in cosmology within the ratio of our cases (a), (b) and (c) has minimal effect.}

\section{Summary}\label{sec:summary}

In this paper we have combined the halo model reaction for modified gravity and non-constant dark energy \citep{Cataneo:2018cic} with the halo model reaction for massive neutrinos \citep{Cataneo:2019fjp}. Combined with a baryonic feedback model and an accurate pseudo spectrum prescription, this offers an analytic means to model theoretically general matter power spectra in the non-linear regime of structure formation at percent level accuracy. We have implemented this extension into the {\tt ReACT} code \citep{Bose:2020wch}.

We have tested the halo model reaction applied to a pseudo spectrum given by the halo model-based prescription of \cite{Mead:2020vgs} ({\tt HMCode2020}) against the DUSTGRAIN-\textit{pathfinder} simulations \citep{Giocoli:2018gqh}, the BAHAMAS hydrodynamical simulations \citep{McCarthy:2016mry,Pfeifer:2020jct}, the Bacco emulator \citep{2020arXiv200406245A} and the official Euclid emulator \citep{Knabenhans:2020gdo}. Our results are summarised in \autoref{resulttab}. 

We find that the theoretical model is generally applicable at the $3\%$-accuracy level for $k\leq 1h/{\rm Mpc}$ for $M_\nu \leq 0.15$eV, $|f_{\rm R0}| \leq 10^{-5}$ and observationally-permitted values of $\{w_0,w_a\}$ (see \cite{Pfeifer:2020jct}). Tests with more accurate pseudo prescriptions, such as COmoving Lagrangian Acceleration (COLA) simulations, show that most of the inaccuracy over this range of scales comes from the {\tt HMCode2020} pseudo spectrum prescription. This was also shown in C19. Improvements to the non-linear pseudo power spectrum are thus essential for achieving the theoretical accuracy requirements of upcoming surveys such as Euclid. 

For low neutrino masses ($M_\nu \leq 0.1$eV) and low deviations to $\Lambda$CDM (e.g. for an $f(R)$ modification, $|f_{\rm R0}| \leq 10^{-6}$), the halo model reaction and {\tt HMCode2020} can be reliably applied at $2\%$ accuracy at scales $k\leq 3h/{\rm Mpc}$. Further, we have also found that evolving dark energy and massive neutrino effects can be reliably modelled independently of baryonic feedback effects, \B{although we aim to test the consistent combination of these effects in future work. This is consistent with the work of \cite{Mummery2017} and \cite{Pfeifer:2020jct} which shows that baryonic effects on the matter power spectrum can be treated independently of the effects of massive neutrinos and evolving dark energy down to an accuracy of $\leq 2\%$ and $\leq 1\%$, respectively. The accuracy of the model proposed in this paper is in superb agreement with current state-of-the-art prescriptions for $\nu w$CDM cosmologies, particularly {\tt HMCode2020}, with the advantage of offering greater model generality and clear pathways for improvement in accuracy.}

On this point, the combination of {\tt HMCode2020} and {\tt ReACT}  offers a very competitive framework to constrain $\nu w$CDM cosmologies \B{as well as modified gravity}, using say cosmic shear data. It can also be used to extend the recent neural network {\tt BaCoN} \citep{Mancarella:2020jyu} to distinguish between non-zero neutrino masses and modified cosmologies and gravity. The modelling of modified gravity theories with scale-dependent growth and massive neutrinos is slightly more restrictive in the scales that it can be reliably applied to. We expect theories with scale-independent  growth, such as the DGP braneworld model \citep{Dvali:2000hr}, to be better modelled by this framework when also including massive neutrinos (see C19 for example).

\begin{table}
\centering
\caption{Maximal percent deviation of the halo model reaction with {\tt HMCode2020} against various benchmarks at different scales when comparing the ratio of the target spectrum to a $\Lambda$CDM spectrum. We show the percent deviation from the benchmark at $z=0$ and $z=1$. We have considered $0.06$eV $\leq M_\nu \leq 0.48$eV, $|f_{\rm R0}| \leq 10^{-5}$ and a broad range of $\{w_0,w_a\}$. \B{EE2 stands for EuclidEmulator2. For the emulator accuracies, we do not consider the `extreme' cases shown in \autoref{fig:ext}.}}
\begin{tabular}{| c | c | c | c | c | }
\hline  
  & \multicolumn{2}{c}{$k \leq 1h/{\rm Mpc}$} & \multicolumn{2}{c}{$k \leq 3h/{\rm Mpc}$} \\ 
 Benchmark & z=0 & z=1  & z=0 & z=1 \\ \hline
 $f(R)+M_\nu$ (DUSTGRAIN-\textit{pathfinder}) & 2\% &3\% & 5\%&4\%   \\ 
 $\nu \Lambda$CDM (BAHAMAS) & 2\%&2\% & 2\%&2\%  \\ 
 $\nu w$CDM (EE2) & 2\% & 1\% & 2\% & 4\%  \\ 
 $\nu w$CDM (Bacco) & 2\%& 1\% & 2\%& 4\%  \\ 
 $\nu w$CDM (BAHAMAS) & 2\%&2\% & 2\%&2\%  \\ 
\end{tabular}
\label{resulttab}
\end{table}

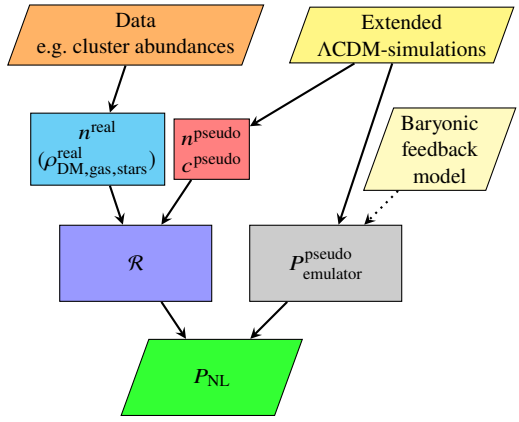
\begin{figure}
\centering
\begin{tikzpicture}[node distance=2cm]
\node (data)  [start , xshift=-3cm, align=center] {Data \\ e.g. cluster abundances};
\node (sims)  [start2 , xshift=0.5cm, align=center] {Extended \\  $\Lambda$CDM-simulations};

\node (pseudosc) [io_halo ,below of=sims,align=center,xshift=-2.5cm, yshift=0.5cm] {$n^{\rm pseudo}$ \\ $c^{\rm pseudo}$};

\node (realsc) [io_halo2 ,below of=data, xshift=-0.5cm, align=center,yshift=0.5cm] {$n^{\rm real}$ \\ ($\rho^{\rm real}_{\rm DM, gas, stars})$};

\node (react) [reaction, below of=realsc,xshift=0.5cm,yshift=0.5cm] {{$\mathcal{R}$}};

\node (pnlpseudo) [io_pseudo ,right of=react,xshift=0.5cm] {$P_{\rm emulator}^{\rm pseudo}$};

\node (pnl) [io_nl ,below of  = react,xshift=1cm,yshift=0.5cm] {$P_{\rm NL}$};

\node (feed) [io_feed ,right of  = pnlpseudo ,xshift=-0.5cm,yshift=1.5cm,align=center] {Baryonic \\ feedback \\ model};

\draw [arrow] (data) -- (realsc);
\draw [arrow] (sims) -- (pseudosc);
\draw [arrow] (sims) -- (pnlpseudo);                  

\draw [arrow] (pseudosc) -- (react);
\draw [arrow] (realsc) -- (react);

\draw [arrow] (react) -- (pnl);
\draw [arrow] (pnlpseudo) -- (pnl);
\draw [arrow,dotted] (feed) -- (pnlpseudo);

\end{tikzpicture}
 \caption[CONVERGENCE]{An idealised pipeline for the non-linear power spectrum (green trapezium) computation. Data sets (orange trapezium), such as cluster abundances, directly constrain the real halo mass function (cyan rectangle) while a suite of $\Lambda$CDM based simulations (yellow trapezium) provide emulated pseudo halo model ingredients (red rectangle) (we exclude the density profile since NFW works sufficiently well). These same simulations are used to construct the pseudo non-linear power spectrum (gray rectangle). The halo model ingredients are provided to a code such as {\tt ReACT} to compute the halo model reaction (blue rectangle). Standard plus the extended cosmological parameters describing beyond-$\Lambda$CDM physics (including neutrino mass) are supplied to both the emulated quantities as well as the reaction computation. The baryonic feedback effects are ideally self consistently included in the halo model through $\rho^{\rm real}_{\rm DM, gas, stars}$ constrained by total matter profiles, or can be optionally added onto $P_{\rm NL}^{\rm pseudo}$ separately through some feedback model (beige trapezium).} 
\label{fig:idealhmr}
\end{figure}

As stated, the main source of inaccuracy for scales $k\leq 3h/{\rm Mpc}$ is the {\tt HMCode2020} pseudo power spectrum prescription.\B{This limiting accuracy is confirmed in our comparisons where we observe sub-percent agreement between the halo model reaction approach and  the `pure' {\tt HMCode2020} predictions for all $\nu w$CDM cosmologies considered.} Further inaccuracy in our approach comes from ill-calibrated mass functions in both pseudo and target cosmologies as well as inaccurate concentration-mass relations. This was also shown in \cite{Srinivasan:2021gib} where they were able to significantly improve the theoretical accuracy by tuning the halo concentration-mass relation within the halo model reaction. Similarly in C19, they show the benefit of using the `correct' $c$-$M$ relation. In \autoref{fig:idealhmr} we illustrate the ideal setup for the non-linear power spectrum predictions under this framework. The pseudo spectrum is given by a bespoke emulator as proposed in \cite{Giblin:2019iit}, as are the ($\Lambda$CDM based) pseudo spherical collapse quantities. The real spherical collapse quantities are parametrised and constrained by astronomical observations, such as cluster abundance data. The real halo density profile used in the reaction can also be constrained by data such as total matter profiles, which would also give a self-consistent model for baryonic effects.

In future work we aim to integrate the massive neutrino section of the code into {\tt CosmoSIS} \citep{Zuntz:2014csq} and perform Markov Chain Monte Carlo analyses on the measured cosmic shear spectrum from beyond-$\Lambda$CDM simulations to set clear scale cuts on the framework as well as forecasts for upcoming surveys. \B{We also aim at testing the independency of baryonic effects and modified gravity/dark energy/massive neutrinos.} As mentioned above, some work in this direction has been carried out by \cite{Mummery2017} and \cite{Pfeifer:2020jct} (for massive neutrino and DE cosmologies), and by \cite{Arnold:2019} (for $f(R)$ gravity). Most likely Vainshtein screening is highly effective at screening halos, so one would expect almost perfect decoupling between baryonic feedback and MG physics.

We are also currently working on generalising the parametrisation of modified gravity and dark energy within the halo model reaction. This could then be validated against parametrised $N$-body simulations which have seen significant development recently \citep{Hassani:2020rxd,Srinivasan:2021gib}.

\section*{Acknowledgments}
\noindent \B{We thank Raul Angulo for sharing the most recent version of the Bacco emulator and helping with the associated comparisons.} We thank Catherine Heymans for useful discussions. \B{We thank Hans Winther for useful discussions about our COLA simulations.} BB and LL acknowledge support from the Swiss National Science Foundation (SNSF) Professorship grant No.~170547. BSW is supported by the Royal Society grant number RGF\textbackslash{}EA\textbackslash{}181023. This project has received funding from the European Research Council (ERC) under the European Union's Horizon 2020 research and innovation programme (grant agreement No 769130).  AP is a UK Research and Innovation Future Leaders Fellow, grant MR/S016066/1. MC and QX acknowledge support from the European Research Council under grant number 647112. 
CG and MB acknowledge the grants ASI n.I/023/12/0, ASI-INAF n.  2018-23-HH.0, 
PRIN MIUR 2015 Cosmology and Fundamental Physics: illuminating the
Dark Universe with Euclid". CG is also supported by the PRIN-MIUR 2017 WSCC32 ``Zooming into dark matter and proto-galaxies with massive
lensing clusters'', PRIN-INAF 2019 "Linking Active Galaxies to Large-Scale Structure: a dataset-oriented approach"
and thankful to the Italian Ministry of Foreign Affairs and International Cooperation, Directorate General
for Country Promotion.
MB also acknowledges support by the project “Combining Cosmic Microwave Background and Large Scale Structure data: an Integrated Approach for Addressing Fundamental Questions in Cosmology”, funded by the PRIN-MIUR 2017 grant 2017YJYZAH. SP acknowledges support from the Deutsche Forschungs Gemeinschaft joint Polish-German research project  LI 2015/7-1.
The DUSTGRAIN-\textit{pathfinder} simulations discussed in this work have been performed and analysed on the Marconi supercomputing machine at Cineca thanks to the PRACE project SIMCODE1 (grantnr. 2016153604, P.I. M. Baldi) and at the Computational Center for Particle and Astrophysics (C2PAP) at the Leibniz Supercomputer Center (LRZ) under the projectID pr94ji. The COLA simulations used in this work were performed on the Sciama High Performance Compute (HPC) cluster which is supported by the ICG, SEPNet, and the University of Portsmouth. This research also utilised Queen Mary’s Apocrita HPC facility, supported by QMUL Research-IT \url{http://doi.org/10.5281/zenodo.438045}. We acknowledge the use of open source software \citep{scipy:2001, Hunter:2007,  mckinney-proc-scipy-2010, numpy:2011}.
\appendix 


\section*{Data Availability}

The software used in this article is publicly available in the {\tt ReACT} repository at \url{https://github.com/nebblu/ReACT/tree/react_with_neutrinos}. The BAHAMAS simulation data used in this work is publicly available at \url{http://powerlib.strw.leidenuniv.nl/#data}. 


\bibliographystyle{mnras}
\bibliography{mybib}

\begin{thebibliography}{}
\makeatletter
\relax
\def\mn@urlcharsother{\let\do\@makeother \do\$\do\&\do\#\do\^\do\_\do\%\do\~}
\def\mn@doi{\begingroup\mn@urlcharsother \@ifnextchar [ {\mn@doi@}
  {\mn@doi@[]}}
\def\mn@doi@[#1]#2{\def\@tempa{#1}\ifx\@tempa\@empty \href
  {http://dx.doi.org/#2} {doi:#2}\else \href {http://dx.doi.org/#2} {#1}\fi
  \endgroup}
\def\mn@eprint#1#2{\mn@eprint@#1:#2::\@nil}
\def\mn@eprint@arXiv#1{\href {http://arxiv.org/abs/#1} {{\tt arXiv:#1}}}
\def\mn@eprint@dblp#1{\href {http://dblp.uni-trier.de/rec/bibtex/#1.xml}
  {dblp:#1}}
\def\mn@eprint@#1:#2:#3:#4\@nil{\def\@tempa {#1}\def\@tempb {#2}\def\@tempc
  {#3}\ifx \@tempc \@empty \let \@tempc \@tempb \let \@tempb \@tempa \fi \ifx
  \@tempb \@empty \def\@tempb {arXiv}\fi \@ifundefined
  {mn@eprint@\@tempb}{\@tempb:\@tempc}{\expandafter \expandafter \csname
  mn@eprint@\@tempb\endcsname \expandafter{\@tempc}}}

\bibitem[\protect\citeauthoryear{Abbott et~al.}{Abbott
  et~al.}{2020}]{Abbott:2020knk}
Abbott T. M.~C.,  et~al., 2020, \mn@doi [Phys. Rev. D]
  {10.1103/PhysRevD.102.023509}, 102, 023509

\bibitem[\protect\citeauthoryear{Agarwal \& Feldman}{Agarwal \&
  Feldman}{2011}]{Agarwal:2010mt}
Agarwal S.,  Feldman H.~A.,  2011, \mn@doi [Mon. Not. Roy. Astron. Soc.]
  {10.1111/j.1365-2966.2010.17546.x}, 410, 1647

\bibitem[\protect\citeauthoryear{Aghanim et~al.}{Aghanim
  et~al.}{2020}]{Aghanim:2018eyx}
Aghanim N.,  et~al., 2020, \mn@doi [Astron. Astrophys.]
  {10.1051/0004-6361/201833910}, 641, A6

\bibitem[\protect\citeauthoryear{{Akeson} et~al.,}{{Akeson}
  et~al.}{2019}]{2019arXiv190205569A}
{Akeson} R.,  et~al., 2019, arXiv e-prints, \href
  {https://ui.adsabs.harvard.edu/abs/2019arXiv190205569A} {p. arXiv:1902.05569}

\bibitem[\protect\citeauthoryear{Amendola et~al.}{Amendola
  et~al.}{2018}]{Amendola:2016saw}
Amendola L.,  et~al., 2018, \mn@doi [Living Rev. Rel.]
  {10.1007/s41114-017-0010-3}, 21, 2

\bibitem[\protect\citeauthoryear{Anderson et~al.}{Anderson
  et~al.}{2013}]{Anderson:2012sa}
Anderson L.,  et~al., 2013, \mn@doi [Mon. Not. Roy. Astron. Soc.]
  {10.1111/j.1365-2966.2012.22066.x}, 427, 3435

\bibitem[\protect\citeauthoryear{{Angulo}, {Zennaro}, {Contreras}, {Aric{\`o}},
  {Pellejero-Iba{\~n}ez}  \& {St{\"u}cker}}{{Angulo}
  et~al.}{2020}]{2020arXiv200406245A}
{Angulo} R.~E.,  {Zennaro} M.,  {Contreras} S.,  {Aric{\`o}} G.,
  {Pellejero-Iba{\~n}ez} M.,   {St{\"u}cker} J.,  2020, arXiv e-prints, \href
  {https://ui.adsabs.harvard.edu/abs/2020arXiv200406245A} {p. arXiv:2004.06245}

\bibitem[\protect\citeauthoryear{{Aric{\`o}}, {Angulo}, {Contreras},
  {Ondaro-Mallea}, {Pellejero-Iba{\~n}ez}  \& {Zennaro}}{{Aric{\`o}}
  et~al.}{2020}]{Arico:2020lhq}
{Aric{\`o}} G.,  {Angulo} R.~E.,  {Contreras} S.,  {Ondaro-Mallea} L.,
  {Pellejero-Iba{\~n}ez} M.,   {Zennaro} M.,  2020, arXiv e-prints, \href
  {https://ui.adsabs.harvard.edu/abs/2020arXiv201115018A} {p. arXiv:2011.15018}

\bibitem[\protect\citeauthoryear{{Arnold} \& {Li}}{{Arnold} \&
  {Li}}{2019}]{Arnold:2019}
{Arnold} C.,  {Li} B.,  2019, \mn@doi [\mnras] {10.1093/mnras/stz2690}, \href
  {https://ui.adsabs.harvard.edu/abs/2019MNRAS.490.2507A} {490, 2507}

\bibitem[\protect\citeauthoryear{Beutler et~al.}{Beutler
  et~al.}{2017}]{Beutler:2016arn}
Beutler F.,  et~al., 2017, \mn@doi [Mon. Not. Roy. Astron. Soc.]
  {10.1093/mnras/stw3298}, 466, 2242

\bibitem[\protect\citeauthoryear{{Bird}, {Viel}  \& {Haehnelt}}{{Bird}
  et~al.}{2012}]{2012MNRAS.420.2551B}
{Bird} S.,  {Viel} M.,   {Haehnelt} M.~G.,  2012, \mn@doi [\mnras]
  {10.1111/j.1365-2966.2011.20222.x}, \href
  {https://ui.adsabs.harvard.edu/abs/2012MNRAS.420.2551B} {420, 2551}

\bibitem[\protect\citeauthoryear{{Bird}, {Ali-Ha{\"\i}moud}, {Feng}  \&
  {Liu}}{{Bird} et~al.}{2018}]{2018MNRAS.481.1486B}
{Bird} S.,  {Ali-Ha{\"\i}moud} Y.,  {Feng} Y.,   {Liu} J.,  2018, \mn@doi
  [\mnras] {10.1093/mnras/sty2376}, \href
  {https://ui.adsabs.harvard.edu/abs/2018MNRAS.481.1486B} {481, 1486}

\bibitem[\protect\citeauthoryear{Blanchard et~al.}{Blanchard
  et~al.}{2020}]{Blanchard:2019oqi}
Blanchard A.,  et~al., 2020, \mn@doi [Astron. Astrophys.]
  {10.1051/0004-6361/202038071}, 642, A191

\bibitem[\protect\citeauthoryear{{Blas}, {Garny}, {Konstandin}  \&
  {Lesgourgues}}{{Blas} et~al.}{2014}]{2014JCAP...11..039B}
{Blas} D.,  {Garny} M.,  {Konstandin} T.,   {Lesgourgues} J.,  2014, \mn@doi
  [\jcap] {10.1088/1475-7516/2014/11/039}, \href
  {https://ui.adsabs.harvard.edu/abs/2014JCAP...11..039B} {2014, 039}

\bibitem[\protect\citeauthoryear{Bose \& Koyama}{Bose \&
  Koyama}{2016}]{Bose:2016qun}
Bose B.,  Koyama K.,  2016, \mn@doi [JCAP] {10.1088/1475-7516/2016/08/032},
  1608, 032

\bibitem[\protect\citeauthoryear{Bose, Cataneo, Tr\"oster, Xia, Heymans  \&
  Lombriser}{Bose et~al.}{2020}]{Bose:2020wch}
Bose B.,  Cataneo M.,  Tr\"oster T.,  Xia Q.,  Heymans C.,   Lombriser L.,
  2020, \mn@doi [Mon. Not. Roy. Astron. Soc.] {10.1093/mnras/staa2696}, 498,
  4650

\bibitem[\protect\citeauthoryear{{Boyle}, {Uhlemann}, {Friedrich},
  {Barthelemy}, {Codis}, {Bernardeau}, {Giocoli}  \& {Baldi}}{{Boyle}
  et~al.}{2020}]{boyle20}
{Boyle} A.,  {Uhlemann} C.,  {Friedrich} O.,  {Barthelemy} A.,  {Codis} S.,
  {Bernardeau} F.,  {Giocoli} C.,   {Baldi} M.,  2020, arXiv e-prints, \href
  {https://ui.adsabs.harvard.edu/abs/2020arXiv201207771B} {p. arXiv:2012.07771}

\bibitem[\protect\citeauthoryear{Bullock, Kolatt, Sigad, Somerville, Kravtsov,
  Klypin, Primack  \& Dekel}{Bullock et~al.}{2001}]{Bullock:1999he}
Bullock J.~S.,  Kolatt T.~S.,  Sigad Y.,  Somerville R.~S.,  Kravtsov A.~V.,
  Klypin A.~A.,  Primack J.~R.,   Dekel A.,  2001, \mn@doi [Mon. Not. Roy.
  Astron. Soc.] {10.1046/j.1365-8711.2001.04068.x}, 321, 559

\bibitem[\protect\citeauthoryear{Cacciato, Bosch, More, Li, Mo  \&
  Yang}{Cacciato et~al.}{2009}]{Cacciato:2008hm}
Cacciato M.,  Bosch F. C. v.~d.,  More S.,  Li R.,  Mo H.~J.,   Yang X.,  2009,
  \mn@doi [Mon. Not. Roy. Astron. Soc.] {10.1111/j.1365-2966.2008.14362.x},
  394, 929

\bibitem[\protect\citeauthoryear{Cataneo, Lombriser, Heymans, Mead, Barreira,
  Bose  \& Li}{Cataneo et~al.}{2019}]{Cataneo:2018cic}
Cataneo M.,  Lombriser L.,  Heymans C.,  Mead A.,  Barreira A.,  Bose S.,   Li
  B.,  2019, \mn@doi [Mon.\ Not.\ Roy.\ Astron.\ Soc.] {10.1093/mnras/stz1836},
  488, 2121

\bibitem[\protect\citeauthoryear{Cataneo, Emberson, Inman, Harnois-Deraps  \&
  Heymans}{Cataneo et~al.}{2020}]{Cataneo:2019fjp}
Cataneo M.,  Emberson J.,  Inman D.,  Harnois-Deraps J.,   Heymans C.,  2020,
  \mn@doi [Mon.\ Not.\ Roy.\ Astron.\ Soc.] {10.1093/mnras/stz3189}, 491, 3101

\bibitem[\protect\citeauthoryear{Chevallier \& Polarski}{Chevallier \&
  Polarski}{2001}]{Chevallier:2000qy}
Chevallier M.,  Polarski D.,  2001, \mn@doi [Int. J. Mod. Phys.]
  {10.1142/S0218271801000822}, D10, 213

\bibitem[\protect\citeauthoryear{Chisari et~al.}{Chisari
  et~al.}{2019}]{Chisari:2019tus}
Chisari N.~E.,  et~al., 2019, \mn@doi [Open J. Astrophys.]
  {10.21105/astro.1905.06082}, 2, 4

\bibitem[\protect\citeauthoryear{Clifton, Ferreira, Padilla  \&
  Skordis}{Clifton et~al.}{2012}]{Clifton:2011jh}
Clifton T.,  Ferreira P.~G.,  Padilla A.,   Skordis C.,  2012, \mn@doi [Phys.
  Rept.] {10.1016/j.physrep.2012.01.001}, 513, 1

\bibitem[\protect\citeauthoryear{{Contarini}, {Marulli}, {Moscardini},
  {Veropalumbo}, {Giocoli}  \& {Baldi}}{{Contarini} et~al.}{2021}]{contarini21}
{Contarini} S.,  {Marulli} F.,  {Moscardini} L.,  {Veropalumbo} A.,  {Giocoli}
  C.,   {Baldi} M.,  2021, \mn@doi [\mnras] {10.1093/mnras/stab1112}, \href
  {https://ui.adsabs.harvard.edu/abs/2021MNRAS.tmp.1085C} {}

\bibitem[\protect\citeauthoryear{Contreras, Angulo, Zennaro, Aric\`o  \&
  Pellejero-Iba\~nez}{Contreras et~al.}{2020}]{Contreras:2020kbv}
Contreras S.,  Angulo R.~E.,  Zennaro M.,  Aric\`o G.,   Pellejero-Iba\~nez M.,
   2020, \mn@doi [Mon. Not. Roy. Astron. Soc.] {10.1093/mnras/staa3117}, 499,
  4905

\bibitem[\protect\citeauthoryear{Cooray \& Sheth}{Cooray \&
  Sheth}{2002}]{Cooray:2002dia}
Cooray A.,  Sheth R.~K.,  2002, \mn@doi [Phys. Rept.]
  {10.1016/S0370-1573(02)00276-4}, 372, 1

\bibitem[\protect\citeauthoryear{Copeland, Sami  \& Tsujikawa}{Copeland
  et~al.}{2006}]{Copeland:2006wr}
Copeland E.~J.,  Sami M.,   Tsujikawa S.,  2006, \mn@doi [Int. J. Mod. Phys.]
  {10.1142/S021827180600942X}, D15, 1753

\bibitem[\protect\citeauthoryear{{Corasaniti}, {Giocoli}  \&
  {Baldi}}{{Corasaniti} et~al.}{2020}]{corasaniti20}
{Corasaniti} P.~S.,  {Giocoli} C.,   {Baldi} M.,  2020, \mn@doi [\prd]
  {10.1103/PhysRevD.102.043501}, \href
  {https://ui.adsabs.harvard.edu/abs/2020PhRvD.102d3501C} {102, 043501}

\bibitem[\protect\citeauthoryear{{Costanzi}, {Villaescusa-Navarro}, {Viel},
  {Xia}, {Borgani}, {Castorina}  \& {Sefusatti}}{{Costanzi}
  et~al.}{2013}]{Costanzi:2013}
{Costanzi} M.,  {Villaescusa-Navarro} F.,  {Viel} M.,  {Xia} J.-Q.,  {Borgani}
  S.,  {Castorina} E.,   {Sefusatti} E.,  2013, \mn@doi [\jcap]
  {10.1088/1475-7516/2013/12/012}, \href
  {https://ui.adsabs.harvard.edu/abs/2013JCAP...12..012C} {2013, 012}

\bibitem[\protect\citeauthoryear{{DeRose} et~al.,}{{DeRose}
  et~al.}{2019}]{DeRose:2019}
{DeRose} J.,  et~al., 2019, \mn@doi [\apj] {10.3847/1538-4357/ab1085}, \href
  {https://ui.adsabs.harvard.edu/abs/2019ApJ...875...69D} {875, 69}

\bibitem[\protect\citeauthoryear{Dolag, Bartelmann, Perrotta, Baccigalupi,
  Moscardini, Meneghetti  \& Tormen}{Dolag et~al.}{2004}]{Dolag:2003ui}
Dolag K.,  Bartelmann M.,  Perrotta F.,  Baccigalupi C.,  Moscardini L.,
  Meneghetti M.,   Tormen G.,  2004, \mn@doi [Astron. Astrophys.]
  {10.1051/0004-6361:20031757}, 416, 853

\bibitem[\protect\citeauthoryear{Dvali, Gabadadze  \& Porrati}{Dvali
  et~al.}{2000}]{Dvali:2000hr}
Dvali G.,  Gabadadze G.,   Porrati M.,  2000, \mn@doi [Phys.Lett.]
  {10.1016/S0370-2693(00)00669-9}, B485, 208

\bibitem[\protect\citeauthoryear{{Euclid Collaboration} et~al.,}{{Euclid
  Collaboration} et~al.}{2020}]{Knabenhans:2020gdo}
{Euclid Collaboration} et~al., 2020, arXiv e-prints, \href
  {https://ui.adsabs.harvard.edu/abs/2020arXiv201011288E} {p. arXiv:2010.11288}

\bibitem[\protect\citeauthoryear{Ferreira}{Ferreira}{2019}]{Ferreira:2019xrr}
Ferreira P.~G.,  2019, \mn@doi [Ann. Rev. Astron. Astrophys.]
  {10.1146/annurev-astro-091918-104423}, 57, 335

\bibitem[\protect\citeauthoryear{{Garc{\'\i}a-Farieta}, {Marulli},
  {Veropalumbo}, {Moscardini}, {Casas-Miranda}, {Giocoli}  \&
  {Baldi}}{{Garc{\'\i}a-Farieta} et~al.}{2019}]{garciaf19}
{Garc{\'\i}a-Farieta} J.~E.,  {Marulli} F.,  {Veropalumbo} A.,  {Moscardini}
  L.,  {Casas-Miranda} R.~A.,  {Giocoli} C.,   {Baldi} M.,  2019, \mn@doi
  [\mnras] {10.1093/mnras/stz1850}, \href
  {https://ui.adsabs.harvard.edu/abs/2019MNRAS.488.1987G} {488, 1987}

\bibitem[\protect\citeauthoryear{Giblin, Cataneo, Moews  \& Heymans}{Giblin
  et~al.}{2019}]{Giblin:2019iit}
Giblin B.,  Cataneo M.,  Moews B.,   Heymans C.,  2019, \mn@doi [Mon. Not. Roy.
  Astron. Soc.] {10.1093/mnras/stz2659}, 490, 4826

\bibitem[\protect\citeauthoryear{Giocoli, Bartelmann, Sheth  \&
  Cacciato}{Giocoli et~al.}{2010}]{Giocoli:2010dm}
Giocoli C.,  Bartelmann M.,  Sheth R.~K.,   Cacciato M.,  2010, \mn@doi [Mon.
  Not. Roy. Astron. Soc.] {10.1111/j.1365-2966.2010.17108.x}, 408, 300

\bibitem[\protect\citeauthoryear{{Giocoli}, {Metcalf}, {Baldi}, {Meneghetti},
  {Moscardini}  \& {Petkova}}{{Giocoli} et~al.}{2015}]{giocoli15}
{Giocoli} C.,  {Metcalf} R.~B.,  {Baldi} M.,  {Meneghetti} M.,  {Moscardini}
  L.,   {Petkova} M.,  2015, \mn@doi [\mnras] {10.1093/mnras/stv1473}, \href
  {https://ui.adsabs.harvard.edu/abs/2015MNRAS.452.2757G} {452, 2757}

\bibitem[\protect\citeauthoryear{Giocoli, Baldi  \& Moscardini}{Giocoli
  et~al.}{2018}]{Giocoli:2018gqh}
Giocoli C.,  Baldi M.,   Moscardini L.,  2018, \mn@doi [Mon. Not. Roy. Astron.
  Soc.] {10.1093/mnras/sty2465}, 481, 2813

\bibitem[\protect\citeauthoryear{{Girelli}, {Pozzetti}, {Bolzonella},
  {Giocoli}, {Marulli}  \& {Baldi}}{{Girelli} et~al.}{2020}]{girelli20}
{Girelli} G.,  {Pozzetti} L.,  {Bolzonella} M.,  {Giocoli} C.,  {Marulli} F.,
  {Baldi} M.,  2020, \mn@doi [\aap] {10.1051/0004-6361/201936329}, \href
  {https://ui.adsabs.harvard.edu/abs/2020A&A...634A.135G} {634, A135}

\bibitem[\protect\citeauthoryear{{Hagstotz}, {Costanzi}, {Baldi}  \&
  {Weller}}{{Hagstotz} et~al.}{2019}]{Hagstotz:2019}
{Hagstotz} S.,  {Costanzi} M.,  {Baldi} M.,   {Weller} J.,  2019, \mn@doi
  [\mnras] {10.1093/mnras/stz1051}, \href
  {https://ui.adsabs.harvard.edu/abs/2019MNRAS.486.3927H} {486, 3927}

\bibitem[\protect\citeauthoryear{Hassani \& Lombriser}{Hassani \&
  Lombriser}{2020}]{Hassani:2020rxd}
Hassani F.,  Lombriser L.,  2020, \mn@doi [Mon. Not. Roy. Astron. Soc.]
  {10.1093/mnras/staa2083}, 497, 1885

\bibitem[\protect\citeauthoryear{{Heymans} et~al.,}{{Heymans}
  et~al.}{2021}]{2021A&A...646A.140H}
{Heymans} C.,  et~al., 2021, \mn@doi [\aap] {10.1051/0004-6361/202039063},
  \href {https://ui.adsabs.harvard.edu/abs/2021A&A...646A.140H} {646, A140}

\bibitem[\protect\citeauthoryear{{Hilbert} et~al.,}{{Hilbert}
  et~al.}{2020}]{hilbert20}
{Hilbert} S.,  et~al., 2020, \mn@doi [\mnras] {10.1093/mnras/staa281}, \href
  {https://ui.adsabs.harvard.edu/abs/2020MNRAS.493..305H} {493, 305}

\bibitem[\protect\citeauthoryear{Hildebrandt et~al.}{Hildebrandt
  et~al.}{2017}]{Hildebrandt:2016iqg}
Hildebrandt H.,  et~al., 2017, \mn@doi [Mon. Not. Roy. Astron. Soc.]
  {10.1093/mnras/stw2805}, 465, 1454

\bibitem[\protect\citeauthoryear{{Hojjati}, {Pogosian}  \& {Zhao}}{{Hojjati}
  et~al.}{2011}]{2011JCAP...08..005H}
{Hojjati} A.,  {Pogosian} L.,   {Zhao} G.-B.,  2011, \mn@doi [\jcap]
  {10.1088/1475-7516/2011/08/005}, \href
  {https://ui.adsabs.harvard.edu/abs/2011JCAP...08..005H} {2011, 005}

\bibitem[\protect\citeauthoryear{Hu \& Sawicki}{Hu \&
  Sawicki}{2007}]{Hu:2007nk}
Hu W.,  Sawicki I.,  2007, \mn@doi [Phys.Rev.] {10.1103/PhysRevD.76.064004},
  D76, 064004

\bibitem[\protect\citeauthoryear{Hunter}{Hunter}{2007}]{Hunter:2007}
Hunter J.~D.,  2007, Computing In Science \& Engineering, 9, 90

\bibitem[\protect\citeauthoryear{Huterer \& Shafer}{Huterer \&
  Shafer}{2018}]{Huterer:2017buf}
Huterer D.,  Shafer D.~L.,  2018, \mn@doi [Rept. Prog. Phys.]
  {10.1088/1361-6633/aa997e}, 81, 016901

\bibitem[\protect\citeauthoryear{Jones, Oliphant, Peterson  et~al.}{Jones
  et~al.}{01  }]{scipy:2001}
Jones E.,  Oliphant T.,  Peterson P.,   et~al., 2001--, {SciPy}: Open source
  scientific tools for {Python}, \url {http://www.scipy.org/}

\bibitem[\protect\citeauthoryear{Joyce, Lombriser  \& Schmidt}{Joyce
  et~al.}{2016}]{Joyce:2016vqv}
Joyce A.,  Lombriser L.,   Schmidt F.,  2016, \mn@doi [Ann. Rev. Nucl. Part.
  Sci.] {10.1146/annurev-nucl-102115-044553}, 66, 95

\bibitem[\protect\citeauthoryear{Koyama}{Koyama}{2018}]{Koyama:2018som}
Koyama K.,  2018, \mn@doi [Int. J. Mod. Phys.] {10.1142/S0218271818480012},
  D27, 1848001

\bibitem[\protect\citeauthoryear{{LSST Dark Energy Science
  Collaboration}}{{LSST Dark Energy Science
  Collaboration}}{2012}]{Abate:2012za}
{LSST Dark Energy Science Collaboration} 2012, arXiv e-prints, \href
  {https://ui.adsabs.harvard.edu/abs/2012arXiv1211.0310L} {p. arXiv:1211.0310}

\bibitem[\protect\citeauthoryear{{Lawrence} et~al.,}{{Lawrence}
  et~al.}{2017}]{2017ApJ...847...50L}
{Lawrence} E.,  et~al., 2017, \mn@doi [\apj] {10.3847/1538-4357/aa86a9}, \href
  {https://ui.adsabs.harvard.edu/abs/2017ApJ...847...50L} {847, 50}

\bibitem[\protect\citeauthoryear{{Levi} et~al.,}{{Levi}
  et~al.}{2019}]{Levi:2019ggs}
{Levi} M.,  et~al., 2019, in Bulletin of the American Astronomical Society.
  p.~57 (\mn@eprint {arXiv} {1907.10688})

\bibitem[\protect\citeauthoryear{Lewis \& Bridle}{Lewis \&
  Bridle}{2002}]{Lewis:2002ah}
Lewis A.,  Bridle S.,  2002, \mn@doi [Phys. Rev.] {10.1103/PhysRevD.66.103511},
  D66, 103511

\bibitem[\protect\citeauthoryear{Li, Li, Wang  \& Wang}{Li
  et~al.}{2011}]{Li:2011sd}
Li M.,  Li X.-D.,  Wang S.,   Wang Y.,  2011, \mn@doi [Commun. Theor. Phys.]
  {10.1088/0253-6102/56/3/24}, 56, 525

\bibitem[\protect\citeauthoryear{Linder}{Linder}{2003}]{Linder:2002et}
Linder E.~V.,  2003, \mn@doi [Phys. Rev. Lett.]
  {10.1103/PhysRevLett.90.091301}, 90, 091301

\bibitem[\protect\citeauthoryear{{Mancarella}, {Kennedy}, {Bose}  \&
  {Lombriser}}{{Mancarella} et~al.}{2020}]{Mancarella:2020jyu}
{Mancarella} M.,  {Kennedy} J.,  {Bose} B.,   {Lombriser} L.,  2020, arXiv
  e-prints, \href {https://ui.adsabs.harvard.edu/abs/2020arXiv201203992M} {p.
  arXiv:2012.03992}

\bibitem[\protect\citeauthoryear{{Martinelli} et~al.,}{{Martinelli}
  et~al.}{2020}]{Martinelli:2020yto}
{Martinelli} M.,  et~al., 2020, arXiv e-prints, \href
  {https://ui.adsabs.harvard.edu/abs/2020arXiv201012382M} {p. arXiv:2010.12382}

\bibitem[\protect\citeauthoryear{Massara, Villaescusa-Navarro  \& Viel}{Massara
  et~al.}{2014}]{Massara:2014kba}
Massara E.,  Villaescusa-Navarro F.,   Viel M.,  2014, \mn@doi [JCAP]
  {10.1088/1475-7516/2014/12/053}, 12, 053

\bibitem[\protect\citeauthoryear{McCarthy, Schaye, Bird  \& Le~Brun}{McCarthy
  et~al.}{2017}]{McCarthy:2016mry}
McCarthy I.~G.,  Schaye J.,  Bird S.,   Le~Brun A. M.~C.,  2017, \mn@doi [Mon.
  Not. Roy. Astron. Soc.] {10.1093/mnras/stw2792}, 465, 2936

\bibitem[\protect\citeauthoryear{McCarthy, Bird, Schaye, Harnois-Deraps, Font
  \& Van~Waerbeke}{McCarthy et~al.}{2018}]{McCarthy:2017csu}
McCarthy I.~G.,  Bird S.,  Schaye J.,  Harnois-Deraps J.,  Font A.~S.,
  Van~Waerbeke L.,  2018, \mn@doi [Mon. Not. Roy. Astron. Soc.]
  {10.1093/mnras/sty377}, 476, 2999

\bibitem[\protect\citeauthoryear{McKinney}{McKinney}{2010}]{mckinney-proc-scipy-2010}
McKinney W.,  2010, in van~der Walt S.,  Millman J.,  eds, Proceedings of the
  9th Python in Science Conference. pp 51 -- 56

\bibitem[\protect\citeauthoryear{Mead}{Mead}{2017}]{Mead:2016ybv}
Mead A.,  2017, \mn@doi [Mon. Not. Roy. Astron. Soc.] {10.1093/mnras/stw2312},
  464, 1282

\bibitem[\protect\citeauthoryear{{Mead}, {Heymans}, {Lombriser}, {Peacock},
  {Steele}  \& {Winther}}{{Mead} et~al.}{2016a}]{2016MNRAS.459.1468M}
{Mead} A.~J.,  {Heymans} C.,  {Lombriser} L.,  {Peacock} J.~A.,  {Steele}
  O.~I.,   {Winther} H.~A.,  2016a, \mn@doi [\mnras] {10.1093/mnras/stw681},
  \href {https://ui.adsabs.harvard.edu/abs/2016MNRAS.459.1468M} {459, 1468}

\bibitem[\protect\citeauthoryear{Mead, Heymans, Lombriser, Peacock, Steele  \&
  Winther}{Mead et~al.}{2016b}]{Mead:2016zqy}
Mead A.,  Heymans C.,  Lombriser L.,  Peacock J.,  Steele O.,   Winther H.,
  2016b, \mn@doi [Mon. Not. Roy. Astron. Soc.] {10.1093/mnras/stw681}, 459,
  1468

\bibitem[\protect\citeauthoryear{{Mead}, {Brieden}, {Tr{\"o}ster}  \&
  {Heymans}}{{Mead} et~al.}{2021}]{Mead:2020vgs}
{Mead} A.~J.,  {Brieden} S.,  {Tr{\"o}ster} T.,   {Heymans} C.,  2021, \mn@doi
  [\mnras] {10.1093/mnras/stab082}, \href
  {https://ui.adsabs.harvard.edu/abs/2021MNRAS.502.1401M} {502, 1401}

\bibitem[\protect\citeauthoryear{{Merten}, {Giocoli}, {Baldi}, {Meneghetti},
  {Peel}, {Lalande}, {Starck}  \& {Pettorino}}{{Merten}
  et~al.}{2019}]{merten19}
{Merten} J.,  {Giocoli} C.,  {Baldi} M.,  {Meneghetti} M.,  {Peel} A.,
  {Lalande} F.,  {Starck} J.-L.,   {Pettorino} V.,  2019, \mn@doi [\mnras]
  {10.1093/mnras/stz972}, \href
  {https://ui.adsabs.harvard.edu/abs/2019MNRAS.487..104M} {487, 104}

\bibitem[\protect\citeauthoryear{{Mummery}, {McCarthy}, {Bird}  \&
  {Schaye}}{{Mummery} et~al.}{2017}]{Mummery2017}
{Mummery} B.~O.,  {McCarthy} I.~G.,  {Bird} S.,   {Schaye} J.,  2017, \mn@doi
  [\mnras] {10.1093/mnras/stx1469}, \href
  {https://ui.adsabs.harvard.edu/abs/2017MNRAS.471..227M} {471, 227}

\bibitem[\protect\citeauthoryear{Navarro, Frenk  \& White}{Navarro
  et~al.}{1997}]{Navarro:1996gj}
Navarro J.~F.,  Frenk C.~S.,   White S.~D.,  1997, \mn@doi [Astrophys.J.]
  {10.1086/304888}, 490, 493

\bibitem[\protect\citeauthoryear{Noller}{Noller}{2020}]{Noller:2020afd}
Noller J.,  2020, \mn@doi [Phys. Rev. D] {10.1103/PhysRevD.101.063524}, 101,
  063524

\bibitem[\protect\citeauthoryear{{Peel}, {Lalande}, {Starck}, {Pettorino},
  {Merten}, {Giocoli}, {Meneghetti}  \& {Baldi}}{{Peel} et~al.}{2019}]{peel19}
{Peel} A.,  {Lalande} F.,  {Starck} J.-L.,  {Pettorino} V.,  {Merten} J.,
  {Giocoli} C.,  {Meneghetti} M.,   {Baldi} M.,  2019, \mn@doi [\prd]
  {10.1103/PhysRevD.100.023508}, \href
  {https://ui.adsabs.harvard.edu/abs/2019PhRvD.100b3508P} {100, 023508}

\bibitem[\protect\citeauthoryear{Pfeifer, McCarthy, Stafford, Brown, Font,
  Kwan, Salcido  \& Schaye}{Pfeifer et~al.}{2020}]{Pfeifer:2020jct}
Pfeifer S.,  McCarthy I.~G.,  Stafford S.~G.,  Brown S.~T.,  Font A.~S.,  Kwan
  J.,  Salcido J.,   Schaye J.,  2020, \mn@doi [Mon. Not. Roy. Astron. Soc.]
  {10.1093/mnras/staa2240}, 498, 1576

\bibitem[\protect\citeauthoryear{{Puchwein}, {Baldi}  \& {Springel}}{{Puchwein}
  et~al.}{2013}]{puchwein13}
{Puchwein} E.,  {Baldi} M.,   {Springel} V.,  2013, \mn@doi [\mnras]
  {10.1093/mnras/stt1575}, \href
  {https://ui.adsabs.harvard.edu/abs/2013MNRAS.436..348P} {436, 348}

\bibitem[\protect\citeauthoryear{{Rogers}, {Peiris}, {Pontzen}, {Bird}, {Verde}
   \& {Font-Ribera}}{{Rogers} et~al.}{2019}]{Rogers:2019}
{Rogers} K.~K.,  {Peiris} H.~V.,  {Pontzen} A.,  {Bird} S.,  {Verde} L.,
  {Font-Ribera} A.,  2019, \mn@doi [\jcap] {10.1088/1475-7516/2019/02/031},
  \href {https://ui.adsabs.harvard.edu/abs/2019JCAP...02..031R} {2019, 031}

\bibitem[\protect\citeauthoryear{Saito, Takada  \& Taruya}{Saito
  et~al.}{2009}]{Saito:2009ah}
Saito S.,  Takada M.,   Taruya A.,  2009, \mn@doi [Phys. Rev. D]
  {10.1103/PhysRevD.80.083528}, 80, 083528

\bibitem[\protect\citeauthoryear{Schneider, Stoira, Refregier, Weiss,
  Knabenhans, Stadel  \& Teyssier}{Schneider et~al.}{2020a}]{Schneider:2019snl}
Schneider A.,  Stoira N.,  Refregier A.,  Weiss A.~J.,  Knabenhans M.,  Stadel
  J.,   Teyssier R.,  2020a, \mn@doi [JCAP] {10.1088/1475-7516/2020/04/019},
  04, 019

\bibitem[\protect\citeauthoryear{Schneider et~al.,}{Schneider
  et~al.}{2020b}]{Schneider:2019xpf}
Schneider A.,  et~al., 2020b, \mn@doi [JCAP] {10.1088/1475-7516/2020/04/020},
  04, 020

\bibitem[\protect\citeauthoryear{Semboloni, Hoekstra, Schaye, van Daalen  \&
  McCarthy}{Semboloni et~al.}{2011}]{Semboloni:2011fe}
Semboloni E.,  Hoekstra H.,  Schaye J.,  van Daalen M.~P.,   McCarthy I.~J.,
  2011, \mn@doi [Mon. Not. Roy. Astron. Soc.]
  {10.1111/j.1365-2966.2011.19385.x}, 417, 2020

\bibitem[\protect\citeauthoryear{Sheth \& Tormen}{Sheth \&
  Tormen}{1999}]{Sheth:1999mn}
Sheth R.~K.,  Tormen G.,  1999, \mn@doi [Mon. Not. Roy. Astron. Soc.]
  {10.1046/j.1365-8711.1999.02692.x}, 308, 119

\bibitem[\protect\citeauthoryear{Sheth \& Tormen}{Sheth \&
  Tormen}{2002}]{Sheth:2001dp}
Sheth R.~K.,  Tormen G.,  2002, \mn@doi [Mon. Not. Roy. Astron. Soc.]
  {10.1046/j.1365-8711.2002.04950.x}, 329, 61

\bibitem[\protect\citeauthoryear{Song et~al.,}{Song
  et~al.}{2015}]{Song:2015oza}
Song Y.-S.,  et~al., 2015, \mn@doi [Phys. Rev.] {10.1103/PhysRevD.92.043522},
  D92, 043522

\bibitem[\protect\citeauthoryear{{Springel} et~al.,}{{Springel}
  et~al.}{2018}]{Springel2018}
{Springel} V.,  et~al., 2018, \mn@doi [\mnras] {10.1093/mnras/stx3304}, \href
  {https://ui.adsabs.harvard.edu/abs/2018MNRAS.475..676S} {475, 676}

\bibitem[\protect\citeauthoryear{{Srinivasan}, {Thomas}, {Pace}  \&
  {Battye}}{{Srinivasan} et~al.}{2021}]{Srinivasan:2021gib}
{Srinivasan} S.,  {Thomas} D.~B.,  {Pace} F.,   {Battye} R.,  2021, arXiv
  e-prints, \href {https://ui.adsabs.harvard.edu/abs/2021arXiv210305051S} {p.
  arXiv:2103.05051}

\bibitem[\protect\citeauthoryear{Takahashi, Sato, Nishimichi, Taruya  \&
  Oguri}{Takahashi et~al.}{2012}]{Takahashi:2012em}
Takahashi R.,  Sato M.,  Nishimichi T.,  Taruya A.,   Oguri M.,  2012, \mn@doi
  [Astrophys. J.] {10.1088/0004-637X/761/2/152}, 761, 152

\bibitem[\protect\citeauthoryear{Taylor, Kitching  \& McEwen}{Taylor
  et~al.}{2018}]{Taylor:2018nrc}
Taylor P.~L.,  Kitching T.~D.,   McEwen J.~D.,  2018, \mn@doi [Phys. Rev. D]
  {10.1103/PhysRevD.98.043532}, 98, 043532

\bibitem[\protect\citeauthoryear{Tram, Brandbyge, Dakin  \& Hannestad}{Tram
  et~al.}{2019}]{Tram:2018znz}
Tram T.,  Brandbyge J.,  Dakin J.,   Hannestad S.,  2019, \mn@doi [JCAP]
  {10.1088/1475-7516/2019/03/022}, 03, 022

\bibitem[\protect\citeauthoryear{{Tr{\"o}ster} et~al.,}{{Tr{\"o}ster}
  et~al.}{2020}]{Troester:2020}
{Tr{\"o}ster} T.,  et~al., 2020, arXiv e-prints, \href
  {https://ui.adsabs.harvard.edu/abs/2020arXiv201016416T} {p. arXiv:2010.16416}

\bibitem[\protect\citeauthoryear{{Van Der Walt}, {Colbert}  \&
  {Varoquaux}}{{Van Der Walt} et~al.}{2011}]{numpy:2011}
{Van Der Walt} S.,  {Colbert} S.~C.,   {Varoquaux} G.,  2011, preprint, \href
  {https://ui.adsabs.harvard.edu/#abs/2011arXiv1102.1523V} {} (\mn@eprint
  {arXiv} {1102.1523})

\bibitem[\protect\citeauthoryear{Winther, Koyama, Manera, Wright  \&
  Zhao}{Winther et~al.}{2017}]{Winther:2017jof}
Winther H.~A.,  Koyama K.,  Manera M.,  Wright B.~S.,   Zhao G.-B.,  2017,
  \mn@doi [JCAP] {10.1088/1475-7516/2017/08/006}, 08, 006

\bibitem[\protect\citeauthoryear{Wright, Winther  \& Koyama}{Wright
  et~al.}{2017}]{Wright:2017dkw}
Wright B.~S.,  Winther H.~A.,   Koyama K.,  2017, \mn@doi [JCAP]
  {10.1088/1475-7516/2017/10/054}, 10, 054

\bibitem[\protect\citeauthoryear{Wright, Koyama, Winther  \& Zhao}{Wright
  et~al.}{2019}]{Wright:2019qhf}
Wright B.~S.,  Koyama K.,  Winther H.~A.,   Zhao G.-B.,  2019, \mn@doi [JCAP]
  {10.1088/1475-7516/2019/06/040}, 06, 040

\bibitem[\protect\citeauthoryear{{Zhao}, {Pogosian}, {Silvestri}  \&
  {Zylberberg}}{{Zhao} et~al.}{2009}]{2009PhRvD..79h3513Z}
{Zhao} G.-B.,  {Pogosian} L.,  {Silvestri} A.,   {Zylberberg} J.,  2009,
  \mn@doi [\prd] {10.1103/PhysRevD.79.083513}, \href
  {https://ui.adsabs.harvard.edu/abs/2009PhRvD..79h3513Z} {79, 083513}

\bibitem[\protect\citeauthoryear{Zucca, Pogosian, Silvestri  \& Zhao}{Zucca
  et~al.}{2019}]{Zucca:2019xhg}
Zucca A.,  Pogosian L.,  Silvestri A.,   Zhao G.-B.,  2019, \mn@doi [JCAP]
  {10.1088/1475-7516/2019/05/001}, 05, 001

\bibitem[\protect\citeauthoryear{Zuntz et~al.,}{Zuntz
  et~al.}{2015}]{Zuntz:2014csq}
Zuntz J.,  et~al., 2015, \mn@doi [Astron. Comput.]
  {10.1016/j.ascom.2015.05.005}, 12, 45

\bibitem[\protect\citeauthoryear{{van Daalen}, {Schaye}, {Booth}  \& {Dalla
  Vecchia}}{{van Daalen} et~al.}{2011}]{vanDaalen2011}
{van Daalen} M.~P.,  {Schaye} J.,  {Booth} C.~M.,   {Dalla Vecchia} C.,  2011,
  \mn@doi [\mnras] {10.1111/j.1365-2966.2011.18981.x}, \href
  {https://ui.adsabs.harvard.edu/abs/2011MNRAS.415.3649V} {415, 3649}

\bibitem[\protect\citeauthoryear{van Daalen, McCarthy  \& Schaye}{van Daalen
  et~al.}{2020}]{vanDaalen:2019pst}
van Daalen M.~P.,  McCarthy I.~G.,   Schaye J.,  2020, \mn@doi [Mon. Not. Roy.
  Astron. Soc.] {10.1093/mnras/stz3199}, 491, 2424

\makeatother
\end{thebibliography}

\appendix 
\section{Simulating the pseudo spectrum with COLA in $f(R)$ gravity with massive neutrinos}\label{app:cola}
Here we check the accuracy of the {\tt HMCode2020} prescription for the non-linear pseudo power spectrum by running a set of COmoving Lagrangian Acceleration (COLA) simulations in $f(R)$ gravity with massive neutrinos using the \B{approach from \cite{Winther:2017jof} and \cite{Wright:2017dkw} that is implemented in the publicly available COLA code {\tt FML}}. These are approximate simulation methods that make use of 2nd order Lagrangian perturbation theory to trade accuracy on small scales for faster speed overall, while keeping accuracy on large scales. \B{We note that due to the approximate nature of these COLA simulations, we do not expect their pseudo spectra, which are essentially a $\Lambda$CDM simulation with modified initial conditions, to match the accuracy of those from {\tt HMCode2020} which is fit to full $N$-body simulations, for $k\gtrsim1h/{\rm Mpc}$ at $z=0$.} We have selected the medium \B{ and high} deviation from $\Lambda$CDM \B{cases, (b) and (c),}  described in \autoref{sec:resultsfr}, so $|f_{\rm R0}|=10^{-5}$ and $M_\nu=0.1$eV \B{(b) and $M_\nu = 0.15$eV (c).} 

\begin{figure}
\centering
  \includegraphics[width=0.48\textwidth,height=0.25\textwidth]{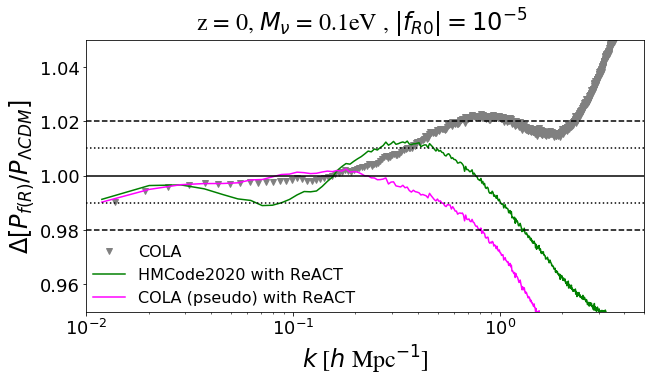}
    \includegraphics[width=0.48\textwidth,height=0.25\textwidth]{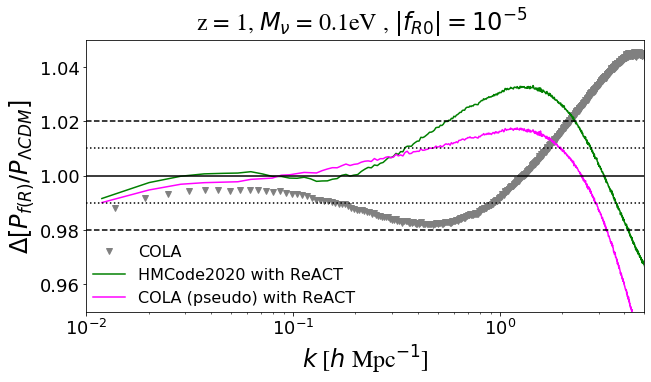}
    \\
  \caption[CONVERGENCE]{Ratio of theoretical predictions to DUSTGRAIN-\textit{pathfinder} measurement of the $f(R)$ to the (massless neutrino) $\Lambda$CDM $P(k)$ ratio with $|f_{\rm R0}|=10^{-5}$ and $M_\nu=0.1$eV. Top is $z=0$ and bottom is $z=1$. We show the full COLA measurements (gray triangles), the {\tt HMCode2020} pseudo with reaction (green solid) predictions and the COLA pseudo with reaction (magenta solid) predictions.}
\label{mnu01f5cola}
\end{figure}

\begin{figure}
\centering
  \includegraphics[width=0.48\textwidth,height=0.25\textwidth]{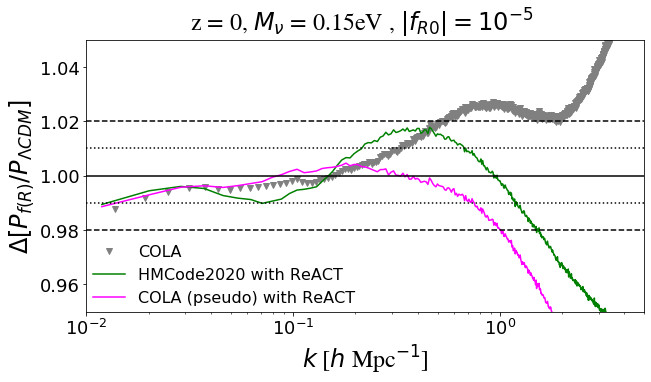}
    \includegraphics[width=0.48\textwidth,height=0.25\textwidth]{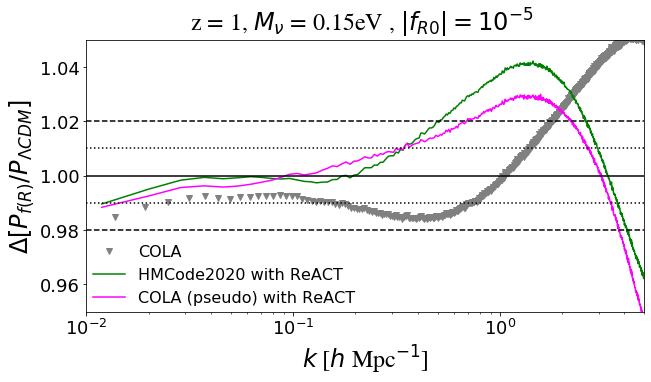}
    \\
  \caption[CONVERGENCE]{Ratio of theoretical predictions to DUSTGRAIN-\textit{pathfinder} measurement of the $f(R)$ to the (massless neutrino) $\Lambda$CDM $P(k)$ ratio with $|f_{\rm R0}|=10^{-5}$ and $M_\nu=0.15$eV. Top is $z=0$ and bottom is $z=1$. We show the full COLA measurements (gray triangles), the {\tt HMCode2020} pseudo with reaction (green solid) predictions and the COLA pseudo with reaction (magenta solid) predictions.}
\label{mnu015f5cola}
\end{figure}

We show the results in \autoref{mnu01f5cola} \B{and \autoref{mnu015f5cola}}. The gray triangles show the ratio of ratios between the two COLA simulation measurements of the power spectrum, in the modified cosmology and $\Lambda$CDM, to the same ratio for the DUSTGRAIN-\textit{pathfinder} $N$-body simulations. This gives an indication of the overall accuracy of the COLA approach. \B{For $M_\nu = 0.1 (0.15) $eV} we find that, at $z=1$, the reaction given in \autoref{eq:reaction} combined with a COLA measured pseudo spectrum is {\it at least} $2 (3)\%$ accurate at scales $k\leq 3h/{\rm Mpc}$. \B{For both cases,} at $z=0$ the $2\%$ accuracy-level of the COLA pseudo with the halo model reaction is guaranteed for scales less than $k \approx 1 h/{\rm Mpc}$, which is roughly the same accuracy as the ratio of the full COLA simulations when compared to DUSTGRAIN-\textit{pathfinder}.

\begin{figure}
\centering
  \includegraphics[width=0.48\textwidth,height=0.25\textwidth]{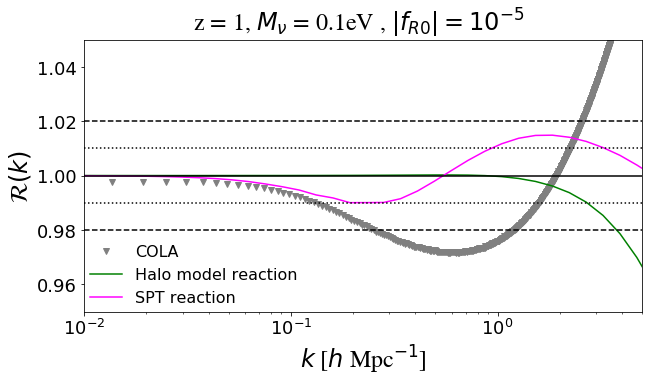}
    \includegraphics[width=0.48\textwidth,height=0.25\textwidth]{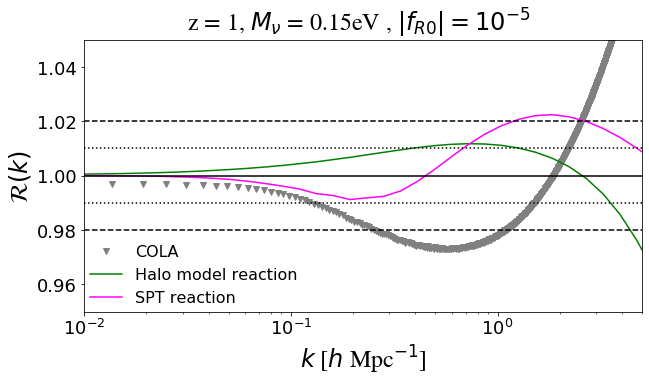}
    \\
  \caption[CONVERGENCE]{The matter power spectrum reaction as measured from COLA (gray triangles), the halo model reaction  prediction (green solid)  and the 1-loop perturbation theory prediction (magenta solid) at $z=1$ for cases (b) (top panel) and (c) (bottom panel).}
\label{reactionfigure}
\end{figure}

Importantly, these comparisons indicate that \B{a significant part of the} inaccuracies seen at $z=1$ at $k\leq 3h/{\rm Mpc}$ for the medium (b) and high (c) deviation cases (see  \autoref{mnu01f5} and \autoref{mnu015f5}) come from the {\tt HMCode2020} pseudo spectrum. \B{We note that further discrepancies, specifically in the $M_\nu=0.15$eV case, come from using inaccurate mass function fits in the  1-halo terms. This is indicated by the enhancement of power the reaction gives the pseudo spectrum (see dotted and solid green lines in the bottom plot of \autoref{mnu015f5}). It is in this case that we get $1 < \mathcal{E} \sim P_{\rm 1h}^{\rm cb} / P_{\rm 1h}^{\rm pseudo}$ which produces this enhancement (see \autoref{eq:reaction} and \autoref{eq:1hcb}). This is highlighted in \autoref{reactionfigure} where we plot the 1-loop perturbation theory prediction for the reaction (see \autoref{eq:sptreaction}), the halo model reaction and the measurement from COLA for $z=1$ in both cases (b) and (c). We clearly see an over-estimation of the halo model reaction at quasi non-linear scales indicating that $\mathcal{E}$ should be less than 1. Note that in these cases we find no solution for $k_\star$ and so $\mathcal{E}$ is set to unity \footnote{The actual value of $\mathcal{E}$ is greater than 1 in both cases.}. We expect this source of inaccuracy to be remedied by measuring the mass function directly from simulations or constraining it using data.}

\bsp	
\label{lastpage}
\end{document}